\documentclass[journal ]{new-aiaa}
\usepackage[utf8]{inputenc}
\usepackage{textcomp}

\usepackage{subcaption}
\usepackage{graphicx}
\usepackage{amsmath}
\usepackage[version=4]{mhchem}
\usepackage{siunitx}
\usepackage{longtable,tabularx}
\usepackage{hyperref}
\title{GPU-based high-precision orbital propagation of large sets of initial conditions through Picard-Chebyshev augmentation}

\author{Alessandro Masat \footnote{PhD Candidate, Department of Aerospace Science and Technology, Via G. La Masa 34, 20156, Milano, Italy, \href{mailto:alessandro.masat@polimi.it}{alessandro.masat@polimi.it}} and Camilla Colombo\footnote{Associate Professor, Department of Aerospace Science and Technology, Via G. La Masa 34, 20156, Milano, Italy, \href{mailto:camilla.colombo@polimi.it}{camilla.colombo@polimi.it}}}
\affil{Politecnico di Milano, Milano, Italy, 20156}
\author{Arnaud Boutonnet \footnote{Senior Mission Analyst, HSO-GFA, ESA-ESOC, Robert-Bosch-Str. 5, D-64293 Darmstadt, Germany, \href{mailto:arnaud.boutonnet@esa.int}{arnaud.boutonnet@esa.int}}}
\affil{European Space Agency (ESA-ESOC), Darmstadt, Germany, 64293}

\begin{document}

\maketitle




\begin{abstract}
The orbital propagation of large sets of initial conditions under high accuracy requirements is currently a bottleneck in the development of space missions, e.g. for planetary protection compliance analyses. The proposed approach can include any force source in the dynamical model through efficient Picard-Chebyshev (PC) numerical simulations. A two-level augmentation of the integration scheme is proposed, to run an arbitrary number of simulations within the same algorithm call, fully exploiting high performance and GPU (Graphics Processing Units) computing facilities. The performances obtained with implementation in C and NVIDIA\textsuperscript{\textregistered} CUDA\textsuperscript{\textregistered} programming languages are shown, on a test case taken from the optimization of a Solar Orbiter-like first resonant phase with Venus.
\end{abstract}
%



\section{Introduction}

Complex trajectory solutions have become a standard choice for interplanetary missions, and often include several gravity assist maneuvers to reach remote space regions with limited fuel consumption. Two recent missions, the ESA/NASA mission Solar Orbiter \cite{Agency2011} and the ESA mission JUICE \cite{JUICE2014}, are meaningful ongoing cases. The former first features preparatory flybys of Earth and Venus, then exploits several resonant encounters with Venus to raise the spacecraft inclination over the ecliptic and observe the Sun's polar regions. The latter requires multiple flybys to reach Jupiter, and then repeatedly swings by different Jupiter's moons to fulfill its scientific observational objectives. Both missions would not be practically feasible without all the designed flybys, as every saved kilogram of fuel means more mass available to board scientific equipment.

The common framework where high-energy multi-flyby trajectories are designed is the patched conics approximation, whose simple but effective model allows a consistent preliminary mission analysis. Feasible gravity assist maneuvers are identified and the consequent optimization of the initial trajectory guess can be performed, by refining the preliminary orbital parameters and/or adding correction maneuvers to meet the operational requirements and fit the full body dynamics.

Nevertheless, the real-life environment features more complex physical phenomena, whose perturbing effects may have significant consequences on the nominal mission. Repeated flybys also imply a higher impact risk of disposal objects with the planets flown by, possibly affecting the mission compliance with planetary protection requirements. COSPAR \cite{COSPAR} maintains worldwide planetary protection policies, e.g. ESA demands the use of at least the N-body Newtonian gravitational model for the compliance assessment \cite{Kminek2012}. This accuracy requirement has so far limited the analysis to Monte-Carlo based techniques \cite{Jehn2014,Wallace2015,Colombo2016,Romano2018,Romano2020} with Cartesian equations of motion, whose simulations inherently carry a heavy computational burden that may limit the mission development time. Recent developments \cite{Masat2022jgcd} nearly halved the required computational time by implementing the Kustaanheimo-Stiefel \cite{Stiefel1971} formulation of the dynamical motion.

This work takes inspiration from the results obtained in \cite{Masat2021AASAIAA,Masat2022acta}, whose aim was to optimize a multi-flyby trajectory arc while taking advantage of the natural dynamics of the N-body relativistic environment. Despite the conceptual complexity, the combined b-plane and Picard-Chebyshev (PC) approach led to an overall computationally feasible algorithm, for a Matlab\textsuperscript{\textregistered} sequential implementation converging to an optimal trajectory in a few minutes only. At the core of the efficiency lies the fixed-point nature of the PC numerical scheme, that minimizes the need of reading and interpolating ephemerides data. The test case is described in more detail in Section \ref{sec:testcase}. The set of trajectories generated by the optimization process is used to validate the concepts proposed in this work, in terms of their parallel propagation with the modified version of the PC method.

The modified PC method has been continuously developed in the past few years, both in its formulation and implementation and outlining possible applications for Earth orbits where it contributed to increase the efficiency of the numerical analyses. Junkins et al \cite{Junkins2013} analyzed the performances of the method comparing the efficiency against the Runge-Kutta-Nystrom 12(10) integrator, proposing also a second order version. Later, Koblick and Shankar \cite{Koblick2015} extended the analysis to the propagation of accurate orbits testing difference force models with NASA's Java Astrodynamics toolkit. Woollands et al. \cite{Woollands2015,Woollands2017,Woollands2018} applied the method as numerical integrator for the solution of the Lambert two-point boundary value problem, assessing also the benefits of adopting the Kuustanheimo-Stiefel formulation of the dynamics and proposing a solution for the multi-revolution trajectory design. Swenson et al. \cite{Swenson2017} applied the modified PC method on the circular restricted three-body problem, using the differential correction approach. Singh et al. \cite{singh2020} used the method as the numerical integration scheme for their feasibility study on quasi-frozen, near polar and low altitude lunar orbits, including the N-bodies and the spherical harmonics perturbations. The fixed point nature of the method was exploited by Koblick et al. \cite{Koblick2016} to design low-thrust trajectories as an optimal control problem, discretizing the control impulses and also included the Earth's oblateness J\textsubscript{2} perturbation. Macomber et al. \cite{macomber2016} introduced the concepts of cold, warm, hot starts of the method, addressing possible efficiency improvements by means of better initial conditions, and variable-precision force models taking advantage of the fixed-point nature of the algorithm. Woollands et al. \cite{Woollands2020}extended the optimal low-thrust design to a high-fidelity model for the non-spherical Earth, considering an arbitrary number of spherical harmonics in the perturbing acceleration. Woollands and Junkins \cite{Woollands2019} developed the Adaptive PC method, including an integral error feedback that accelerates the convergence of the Picard iterations and an empirical law to determine segment length and polynomial degree of the method, based on previous stability analyses. Atallah et al. \cite{Atallah2020} compared the method with other sequential integration techniques on different Earth-based orbital cases.

Despite the parallelization possibilities, the method's instability when dealing with long integration spans and the limited number of nodes (about 200) required to reach machine precision levels are disadvantages for the massively parallel/GPU (Graphics Processing Units) implementation of the PC integration scheme. Differently from the previous developments of the method, this work introduces an augmentation of the dynamical system being integrated to enable stable massive parallelism for the short-term propagation of large sets of initial conditions. The results obtained in \cite{Masat2021AASAIAA,Masat2022acta}, as all the simulated trajectories analyzed to find the optimal one, are re-run as a direct application of the proposed approach, discussing the parallelization options in deep detail. Within a single run, the common time nodes of all the trajectory arcs allow to rework the iterative refinement process of PC. Instead of the six-dimensional Cartesian state of the single trajectory, the propagation is performed for a two-level augmented state, made by a properly sorted collection of the states of multiple trajectories at the same time node. Regardless the sorting choices of the single states within the augmented system, the fundamental mathematical structure of the PC process remains unaltered, without the need of re-defining the method's steps, coefficients and matrices. The integration of the newest developments of the method, particularly the second order version \cite{Junkins2013} with error feedback \cite{Woollands2019}, would therefore be applicable to the augmented version as well. The two different augmentation levels allow to preserve the fine grain flexibility that, if feasible, simulating each single trajectory alone would have, without a too large sacrifice in computational efficiency.

Based on a different approach, GPU-based programs for the propagation of large sets of initial conditions already exist: Geda et al. \cite{Geda2019} developed for the European Space Agency the CUDAjectory tool, a GPU-based propagator that implements the Runge-Kutta-Fehlberg 7/8 scheme. The massive parallelism is exploited by taking the forward integration step in parallel for many different samples, parallelizing the evaluation of the dynamics function and the simple matrix multiplications involved in the evaluation of the intermediate sub steps. Another interesting feature implemented in CUDAjectory is the optimized ephemerides reading and storage technique, following the work of Schrammel et al. \cite{Schrammel2020}, which increases the data locality and minimizes the required memory transactions. Despite also propagating large sets of trajectories, the proposed work differs from the CUDAjectory approach in the adopted numerical scheme. The PC method is by itself a parallelizable routine, the augmented version proposed and implemented in this work contributes to enhance this aspect, resulting in a highly scalable and performing program.

On a final note, the concept of the proposed approach may extend to the more general uncertainty propagation case, applied also to planetary and terrestrial system, for robust analysis or collision probability computation. The heavy computational burden has promoted the adoption of simplified approaches, that aim at achieving acceptable accuracy in the propagated uncertainty for lower computational costs \cite{Li2022}. The heavy burden of accurate Monte Carlo simulations was found to be a major challenge, for instance, even in the design of the touchdown of the Hayabusa 2 mission \cite{Yoshikawa2020} and the characterisation of the impact probability with Earth of the Asteroid Apophis \cite{Armellin2010}. It is beyond the scopes of this work to detail the existing uncertainty propagation techniques that have been developed to reduce the computational burden. The only similarity with the proposed approach is the accessible reduction of the computational time, that is achieved using an inexpensive GPU, whose logic could be transferred to any other uncertainty propagation method that requires the simulations of relatively large sets of initial conditions.


\section{Picard-Chebyshev integration method}\label{sec:PC}

\subsection{Integration concept}
Picard iterations \cite{Hairer1993} are a method that can be used to obtain an approximation of the solution of initial/boundary value problems. Denoting the state of dimension $n$ with $\textbf{x}$, the independent variable with $t$, the initial/boundary condition with $\textbf{x}_0$ and the dynamics function with $\textbf{f}(\textbf{x},t)$, the problem is defined as:
\begin{equation}\label{eq:ibvp}
	\begin{gathered}
		\frac{d \textbf{x}}{d t} = \textbf{f} (\textbf{x},t), \qquad \textbf{x}_0 = \textbf{x}(t_0)
	\end{gathered}
\end{equation}

Starting from an initial approximation $\textbf{x}^{(0)}(t)$ of the actual solution $\textbf{x}(t)$ in the interval $\big[t_0,t\big]$ of the initial/boundary value problem presented in Equation \eqref{eq:ibvp}, the $i$-th Picard iteration improves the previous approximation $\textbf{x}^{(i-1)}(t)$ of $\textbf{x}(t)$ with $\textbf{x}^{(i)}(t)$ as in \cite{Hairer1993}:
\begin{equation}
	\begin{gathered}
		\textbf{x}^{(i)}(t) = \textbf{x}^{(0)}(t) + \int_{t_0}^{t} \textbf{f}\big(\textbf{x}^{(i-1)}(s),s\big) ds
	\end{gathered}
\end{equation}

The method converges for a good enough initial approximation $\textbf{x}^{(0)}(t)$ and for $i \longrightarrow +\infty$ \cite{Hairer1993}.

In the analytical Picard iteration context, performing more than one iteration is in general hard. The increasingly complex expressions for $\textbf{x}^{(i)}(t)$ make it difficult to retrieve closed form solutions after the first 2-3 steps \cite{Bai2010}. At the same time, numerically computing the integral functions by quadrature might not suffice in accuracy, as only the first few iterations in general improve the function approximation. In the attempt to develop parallelizable routines for the integration of the dynamical motion, the PC method was built combining the Picard iterations with the Chebyshev polynomial approximation \cite{Rivlin1976}. A possible derivation of the method that follows the work of Fukushima \cite{Fukushima1997} can be summarized in three steps:
\begin{enumerate}
	\item Select a good enough initial guess $\textbf{x}^{(0)}(t)$.
	\item Approximate $\textbf{f}(\textbf{x},t)$ and $\textbf{x}^{(0)}(t)$ with their Chebyshev polynomial expansion.
	\item Perform a Picard iteration to update the coefficients of the interpolating Chebyshev polynomials.
\end{enumerate}
The Picard iterations halt when the stopping conditions are met, based on the maximum difference between two consecutive iterations dropping below some user-specified tolerance.

The so defined method allows to easily perform several more Picard iterations than the analytical case. The involved expressions remains always of the same type, i.e. the Chebyshev polynomials. The function approximation becomes an interpolation through nodes that should be close to the true trajectory, instead of a global function whose value after the iterations still depend on the initial guess choice. Furthermore, few iterations suffice to drop below a low tolerance if the real solution $\textbf{x}(t)$ differs from the initial guess $\textbf{x}^{(0)}(t)$ only because of small perturbations \cite{Hairer1993}. Starting from the unperturbed Keplerian solution for the generic weakly perturbed two body problem, a relatively fast convergence of the method is ensured \cite{Fukushima1997}. In the context of orbital simulations, Macomber \cite{macomber2015} referred to this type of initial guess as warm-starting the PC iteration method, because the analytic solution of the dominant dynamics part is used to reduce the number of iterations required. Differently, the cold start was defined by simply setting all the trajectory samples as equal to the initial condition. In general, the closer the initial guess to the true trajectory, the lower the number of iterations will be. Semi-analytic initial guesses or results of propagations from simpler models are also an options, and in the case of three-body-like perturbed trajectories would be a better choice compared to the Keplerian approximation. Macomber also introduced the concept of hot start in the case of time spans covering multiple Earth planetary orbits \cite{macomber2015}, where the first orbit was used to compute the difference between the Keplerian guess and the converged trajectory. The near-periodicity of the spherical harmonics perturbation was then exploited, including this difference in the starting trajectory, achieving a further reduction of the iterations required for convergence.

\subsection{Matrix form for vectorized and parallel computation}

The method is suitable for parallel or vector implementation (indeed Fukushi-ma also proposed a vectorized version \cite{Fukushima1997b}), in particular for the evaluation of the dynamics function and the execution of the matrix multiplications. More recent works over this technique by Bai and Junkins developed the modified PC method \cite{Bai2011} and a CUDA implementation for NVIDIA GPUs \cite{Bai2010}. For compactness and to better highlight the parallelization possibilities, the method is presented following the matrix formulation by Koblick et al \cite{Koblick2012}. 

For $N$ Chebyshev nodes and the integration interval $[t_0,t_{N-1}]$, the independent variable $t$ is sampled for $j = 0,1,...,N-1$ up-front as
\begin{equation}\label{eq:chnodes1}
	\begin{aligned}
		t_j & = \omega_2 \tau_j + \omega_1
	\end{aligned}
\end{equation}
with
\begin{equation}\label{eq:chnodes2}
	\begin{aligned}
	    \tau_j & = -\cos\bigg( \frac{j\pi}{N-1} \bigg), &
		\omega_1 & = \frac{t_{N-1}+t_0}{2}, &
		\omega_2 & = \frac{t_{N-1}-t_0}{2} 
	\end{aligned}
\end{equation}

Given the $n$-dimensional sampled states $\textbf{y}^{(i-1)}(t_j) = \textbf{y}^{(i-1)}_j, \enskip j = 0,...,N$ as a matrix $\textbf{y}^{(i-1)}$ of dimension $N \times n$ computed at the Picard iteration $i-1$, the whole process can be summarized in three sequential steps to obtain the states at the iteration $i$. The first one collects the evaluations of the dynamics function $\textbf{f}$ in the $N \times n$ force matrix $\textbf{F}$ \cite{Koblick2012}:
\begin{equation}
	\begin{aligned}
		\textbf{F}^{(i)}_{j+1} = \omega_2 \enskip \textbf{f}\big(\textbf{y}^{(i-1)}_j,t_j\big), \quad & j = 0,...,N-1
	\end{aligned}
\end{equation}

Secondly, identifying with $\textbf{A}$, $\textbf{C}$, $\textbf{S}$ the method's constant matrices whose definition can be found in \cite{Koblick2012}, the $N \times n$ matrix $\textbf{B}$ is obtained by rows as \cite{Koblick2012}
\begin{equation}
	\begin{aligned}
		\textbf{B}_1 & = \textbf{S} \textbf{A} \textbf{F} + 2 \textbf{y}_0, &
		\textbf{B}_{j} & = \textbf{A} \textbf{F},  & j = 2,...,N \\
	\end{aligned}
\end{equation}

Third and last, the $N \times n$ matrix of the state guesses $\textbf{y}^{(i)}$ for the $i$-th Picard iteration is
\begin{equation}
	\begin{aligned}
		\textbf{y}^{(i)} = \textbf{C} \textbf{B}
	\end{aligned}
\end{equation}

The iteration process stops when the maximum state difference between two consecutive Picard iterations $\textbf{y}^{(i)}$ and $\textbf{y}^{(i-1)}$ drops below a specified relative or absolute tolerance, upon user's choice.

Despite the proved theoretical convergence, large integration spans may lead to numerical instabilities, due to the cumulation of round-off errors even with large $N$ as multiple orbital revolutions take place \cite{Fukushima1997,Bai2010,Bai2011}. Fukushima \cite{Fukushima1997} suggests a piece-wise approach as a workaround, which has been implemented in this work and uses the modified PC method to integrate orbit by orbit in sequence\footnote{The proposed implementation automatically handles either forward or backward integration.} until the end of the span. 

The core steps of the proposed algorithm follow the presented scheme \cite{Bai2010,Bai2011}, together with the automatic generation of the Keplerian initial guess spanning one nominal orbital period.

\section{Two-level system augmentation}\label{sec:AugPC}
The just described instability problems make the PC algorithm not efficient for massive parallelism, since increasing the number of nodes would not result in improved accuracy after a certain point. Nevertheless, an increase in parallel efficiency can be found when more trajectories are propagated, in the way the following lines describe.

\subsection{One-level augmentation}\label{subsec:onelevel}
Instead of the evolution of the sole trajectory determined by the initial condition $\textbf{y}_0$, the system being integrated can be re-written so that $M$ different trajectories sampled on the same $N$ time nodes can be processed within a unique iterative process. At the iteration $i$, the matrix $\textbf{Y}^{(i)}$ collects all the samples of all the trajectories, its $j$-th row is related to the $j$-th time sample of the $m$-th trajectory by:
\begin{equation}
	\begin{gathered}
		\textbf{Y}^{(i)}_{j} = \bigg[\textbf{y}^{(i)}_{j,1} \enskip \cdots \enskip \textbf{y}^{(i)}_{j,m} \enskip \cdots \enskip \textbf{y}^{(i)}_{j,M} \bigg], \quad j=1,...,N
	\end{gathered}
\end{equation}
and similarly for the dynamics function evaluations collected in the matrix $\textbf{F}^{(i)}$:
\begin{equation}
	\begin{gathered}
		\textbf{F}^{(i)}_{j} = \bigg[\textbf{F}^{(i)}_{j,1} \enskip \cdots \enskip \textbf{F}^{(i)}_{j,m} \enskip \cdots \enskip \textbf{F}^{(i)}_{j,M} \bigg], \quad j=1,...,N
	\end{gathered}
\end{equation}
whose elements are still computed per sample:
\begin{equation}
	\begin{gathered}
		\textbf{F}^{(i)}_{j,m} = \omega_2 \enskip \textbf{f}\big(\textbf{y}^{(i-1)}_{j,m},t_{j-1}\big), \quad j = 1,...,N
	\end{gathered}
\end{equation}

In principle, building the augmented system only requires to define $\textbf{Y}^{(i)}$ by stacking the different $M$ trajectory matrices along the columns, and similarly $\textbf{F}^{(i)}$ undergoes the exact same modification. The structure of the PC iterations remains unchanged and features the usual steps. First, evaluate the dynamics function for all the $N$ states of all the $M$ trajectories with $\textbf{F}^{(i)}_{j,m} = \omega_2 \enskip \textbf{f}\big(\textbf{y}^{(i-1)}_{j,m},t_{j-1}\big)$. Second, perform the matrix operations $\textbf{B}_1 = \textbf{S} \textbf{A} \textbf{F} + 2 \textbf{Y}_0$ and $\textbf{B}_{j} = \textbf{A} \textbf{F}$, for $j = 2,...,N$. Third and last, update the guesses for all the $M$ trajectories with $\textbf{Y}^{(i)} = \textbf{C} \textbf{B}$.

\subsection{Two-level augmentation}
The stack-along-column rule can be applied again, this time collecting in one single matrix $P$ groups of different $M_{p}$ trajectories each. The augmented matrix $\textbf{Y}^{(i)}$ is now built as
\begin{equation}
	\begin{gathered}
		\textbf{Y}^{(i)}_{j} = \bigg[\textbf{Y}^{(i)}_{j,1} \enskip \cdots \enskip \textbf{Y}^{(i)}_{j,p} \enskip \cdots \enskip \textbf{Y}^{(i)}_{j,P} \bigg], \quad j=1,...,N
	\end{gathered}
\end{equation}
with
\begin{equation}
	\begin{gathered}
		\textbf{Y}^{(i)}_{j,p} = \bigg[\textbf{y}^{(i)}_{j,p,1} \enskip \cdots \enskip \textbf{y}^{(i)}_{j,p,m} \enskip \cdots \enskip \textbf{y}^{(i)}_{j,p,M_{p}} \bigg], \quad j=1,...,N
	\end{gathered}
\end{equation}

In principle, infinite augmentation levels could be built relying on the same logic, none of them would require modifications in the core PC algorithm structure. Nevertheless, a re-definition of the iteration error can be helpful for practical purposes, since the augmentation rationale is purely computational.

Two strategies can be addressed. The first uses a traditional error definition, that treats the trajectory samples as if they were part of a unique system, whose maximum will be compared against the iteration stopping condition. The second introduces a more flexible per-block error definition, that treats the different trajectory blocks as independent, for which the augmentation has then a sole computational purpose. Both the approaches have advantages and drawbacks. The former would allow a simpler implementation and is inevitably computationally more efficient than the latter, because of the reduced overhead compared to maintaining the group split. However, dissimilar trajectories requiring a significantly different number of iterations would keep the computational resources busy for already converged blocks, while the per-block definition allows far more flexibility on this regard.

This work uses the two augmentation levels to build a hybrid approach, that treats the outer blocks as independent, and the inner ones as a single system in a more strict sense. In this way, groups of similar trajectories can be considered as unique but separate augmented system, allowing to maximize the integration performances. The two level augmentation also provides a framework to deal with single trajectories within the same high performance computing context, considering them as a group made of only one member. As a practical example, the outer augmentation level could be used to "isolate" a sub-group of trajectories experiencing a planetary flyby, since their dynamics would become significantly different from the rest of the samples.

\section{Parallel computing, GPU computing and NVIDIA\textsuperscript{\textregistered} CUDA\textsuperscript{\textregistered} fundamentals}\label{sec:GPU}
Parallelizable programs are characterized by some of their parts that can be executed simultaneously. Conceptually, task-parallel and data-parallel routines may exist: the former features completely independent tasks that do not need to be executed one before the other, the latter is identified by a common and repeated task that should be executed on different data. This work focuses on this second aspect, whose features enable efficient GPU computing when massive parallelization is possible. For instance, taking the product of two matrices is a highly parallelizable task, as any element of the result matrix could be computed in parallel. Similarly, in the orbital propagation of large sets of initial conditions\footnote{Assuming they are not interacting with each other and have negligible mass.} the dynamics function could be evaluated in parallel for all the states.

\subsection{Shared memory parallelism with multiple CPUs: OpenMP\textsuperscript{\textregistered}}
Prior to discussing the rationale of GPU computing, the most straightforward parallelization concept involves the use of multiple CPUs (Central Processing Units). Shared memory parallelism is the simplest scenario, where all the machine compute cores have access to the same and common memory locations. More complex supercomputers use however a distributed memory logic, where group of CPUs access their own independent memory. Such systems also include a communication network, to distribute and collect the computed data on the different nodes, and require their own programming paradigm that includes the data and message passing routines \cite{MPI}.

OpenMP is a set of pre-processor instructions that enable simple shared memory parallelization of C, C++ and Fortran programs \cite{openmp}, which take action during the code compilation. Only minor modifications are required to accelerate the most intensive parts of the program, and OpenMP instructions are simply ignored and treated as comments if the OpenMP compilation flag is disabled. Common parallelization instructions involve for loops, which can include some extra functionality: for instance, perfectly nested loops can be collapsed into a single larger loop, increasing the program efficiency, and specific instructions can aid the memory management. Despite the simple interface, most programs require at least some variables to be declared private to each worker, since OpenMP treats all the variables as shared by default. The typical case of variables that should be made private are the loop counters: it is fundamental that each thread works with its own variable, particularly for those cases where the loop counter also identifies the position in a shared array where to access some data. More complex programming scenarios often arise and require the programmer to explicitly control concurrent updates of shared variables, it is beyond the scopes of this work to tackle them all. The reader can refer to the OpenMP programming guide for more complete and detailed information \cite{openmp}. This paragraph serves as a simple introduction to the tool that is used to parallelize the C version of the proposed program.

The parallel dynamics function is implemented as the simple OpenMP for loop parallelization, collapsing all the states of the augmented system into a single loop. The multiple workers access the shared state array, then they compute the acceleration values and temporarily store them into thread-private variables, and finally they copy them back to a shared and global acceleration array. OpenMP exploits the flexibility of the CPU architecture, thus no significant modification are required to the innermost parts of the dynamics function to make an efficiently parallelized program. OpenMP is also used "indirectly" for the matrix multiplications of the PC method. The optimized OpenBLAS \cite{openblas,openblas1,openblas2} libraries are used to implement this part of the program, they already include the OpenMP parallelization. 

\subsection{Parallel reduction}
\begin{figure}[h!]
	\tiny
	\centering 
	\def\svgwidth{\columnwidth}
	\includegraphics[scale=0.75]{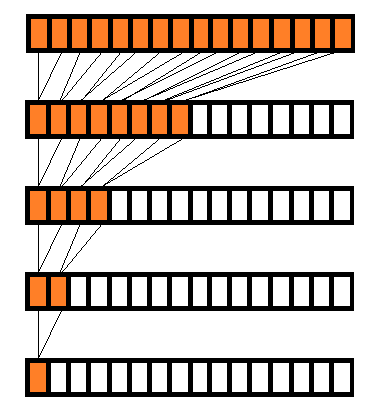}
	\caption{Parallel reduction graphical scheme.}
	\label{fig:reduction}
\end{figure}

The parallelization potential is also exploited in the computation of the PC iteration error, through the parallel reduction mechanism. Reduction tasks are a broad category of compute operations, whose aim is to extract a single scalar value from an array of elements. Some examples are the sum of array elements, finding the maximum or minimum element in an array, "and" and "or" logical operators. Despite appearing intrinsically sequential tasks, parallelization possibilities do exist even in the reduction case: in principle, the whole array is split into several chunks on the parallel workers, which cooperate to perform the reduction task on their own chunk of elements. The cooperated process continues, until a single scalar reduced value is obtained. Figure \ref{fig:reduction} shows a graphical example of the parallel reduction logic: on an array of 16 elements, only 5 sequential steps are eventually required with 16 parallel workers. 
OpenMP implements the reduction clause among its functions, the programmer is only asked to define the final reduced scalar as a shared variable \cite{openmp}. The compiler then ensures that all the array values are scanned through and avoids simultaneous overwriting of the reduced scalar.

\subsection{GPU computing}
\begin{figure}[h!]
	\tiny
	\centering 
	\def\svgwidth{\columnwidth}
	\includegraphics[width=\columnwidth]{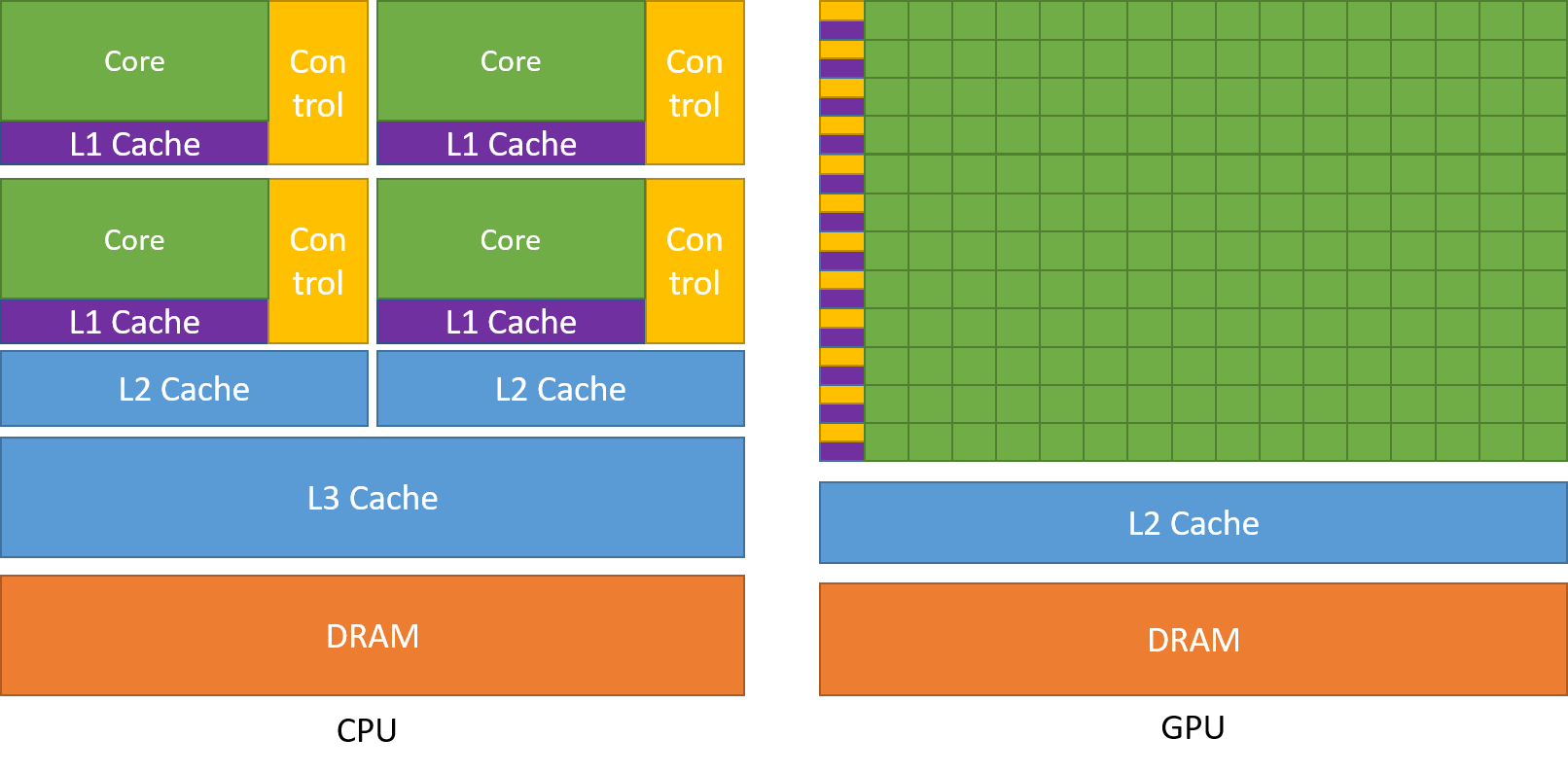}
	\caption{CPU vs GPU architecture difference graphical scheme. Picture from \cite{CUDA}.}
	\label{fig:gpuarch}
\end{figure}

The features of GPU computing arise from the hardware architecture of graphics cards, which is profoundly different from the traditional compute units. Figure \ref{fig:gpuarch} shows a graphical representation of such differences: in summary, more transistors are devoted to pure data processing on GPUs, instead of flow control and memory management as in the CPU case \cite{CUDA}. Corresponding colors refer to chip elements of the same type, i.e. green for compute cores, yellow for control units, violet for the core-level cached memory, blue for shared cache, and orange for global memory.

The processing units are grouped in blocks (typically 32 processing units per block), each controlled by one controller. All processing units in the same block all execute the same instruction, issued only once by the controller. This aspect, together with the normally hundreds to thousands of processing units available in modern graphics cards, makes GPUs prone to implement massive parallelism, although with lower flexibility and higher programming effort compared to CPU applications. Some key concepts are given in the following subsection, a comprehensive view can be found in the CUDA C++ programming guide \cite{CUDA}.

\subsection{Main programming paradigms and the CUDA\textsuperscript{\textregistered} API}
This section is intended to provide a brief overview and nomenclature of the CUDA language and API. Italic font is used to introduce CUDA-specific names and concepts. A complete description can be found in the CUDA user manual \cite{CUDA}.

The fundamental execution unit is called \textit{thread}. Threads can be grouped in \textit{blocks}, and some \textit{shared memory} (dozens of kilobytes) is available to all threads in a common block. The execution of a single instruction is always performed by groups of 32 threads at the same time, called a \textit{warp}, regardless the number of threads in a block. Therefore, blocks with less than 32 threads do not exploit the full hardware resources. Complex configurations can be achieved combining multiple CPUs and GPUs to run the same program. The program flow is always controlled by the CPU, which also controls the execution of the GPU. A function that is invoked by the CPU but executed on the GPU is called \textit{kernel}.

\begin{figure}[h!]
	\tiny
	\centering 
	\def\svgwidth{\columnwidth}
	\includegraphics[scale=0.55]{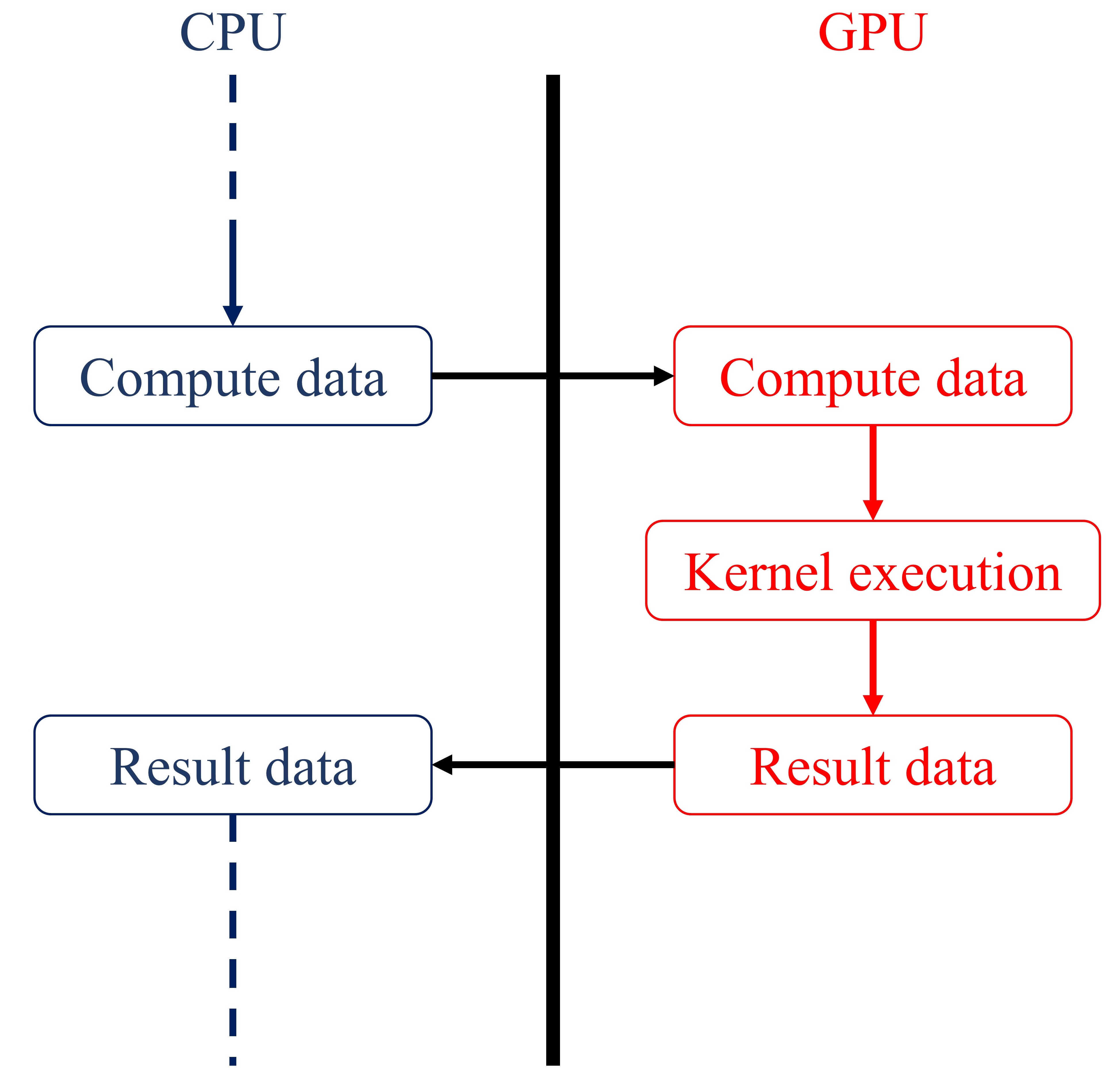}
	\caption{GPU program basic flow.}
	\label{fig:baseflow}
\end{figure}

The first difference compared to standard programming regards the memory access: the GPU cannot read the usual compute memory, but data must be loaded on GPU memory prior to running kernels. In a similar manner, data must be retrieved to the CPU after the kernel has completed its execution, before executing other CPU tasks on the same data. Consequently, the flow of a GPU program/sub-program always follows a first initialization on the host, a subsequent data movement from the CPU to the GPU, the kernel execution, and finally the data retrieval from the GPU to the CPU, as represented in Figure \ref{fig:baseflow}.

Because all threads execute the same instruction at the same time, kernels must be programmed in a warp-oriented manner. This also includes explicitly managing the data access by the various threads, and complex functions typically require the programmer to optimize memory access and cache utilization by hand: while it is implicitly controlled by the compiler for CPUs, graphics card store data by default in the so called \textit{global} memory, large in size (some gigabytes) and referred to by default when GPU variables are initialized. This eases data movements between CPU and GPU, even in case of large arrays. However, the access latency is much higher compared to shared memory: this makes global memory not efficient for repeated read/write operations with GPU variables. To overcome this limitation, shared memory can be exploited for compute purposes to keep read-only values on lower latency locations, instead of only holding common values for all the threads in a block. Few cached bytes are also available on \textit{registers}, thread-private locations to store up to 256 single precision floating point values. Their small size and the compute intensity of the relativistic dynamics function makes it difficult to use registers instead of the shared memory for the proposed application. Their use is limited to temporary and handle variables that aid the final acceleration computation.

The way array elements are sorted is also fundamental for optimized data movements across global and shared GPU memory. While CPU-GPU data transfers are specified by the array size, the most efficient intra-GPU memory access happens when threads read/write values on adjacent locations. If this condition is satisfied, data are moved as a single memory transaction for all the threads, resulting in minimized cycles spent reading or writing on global memory. This memory access pattern is called \textit{coalesced}, and represents a fundamental performance driver in complex GPU programs: even if a program has massive parallelization possibilities, non-optimal memory access may result in the memory access latency not compensated at all by the parallelized computational tasks.

NVIDIA\textsuperscript{\textregistered} developed and maintains the CUDA\textsuperscript{\textregistered} programming language, which remarkably simplifies the use of NVIDIA\textsuperscript{\textregistered} GPUs in computer programs. It is built as a C++ extension, with a set of keywords and API functions that allow programmers to build their own kernels and control the GPU execution flow. A set of optimized libraries is also available, for instance the basic linear algebra cuBLAS\textsuperscript{\textregistered} implemented for the PC matrix multiplications of this work \cite{CUDA}.

\subsection{Concurrency and advanced features} \label{sec:CUDAadvanced}
The CPU-GPU duality and cooperation exposes more possibilities, other than the simple acceleration of intensive parts of the program. In general, kernel calls are asynchronous and do not block the work of the CPU, which allows the CPU to execute other tasks while a GPU kernel is still running. Furthermore, modern GPUs can manage at the same time two  memory transfers (one per direction, CPU to GPU and GPU to CPU) while saturating its compute units for one or more concurrent kernel execution \cite{CUDA}.

CUDA\textsuperscript{\textregistered} allows the queuing of a series of sequentially dependent GPU function calls with the use of \textit{streams}: for instance, an application may require some data to be transferred to the GPU before the execution of a custom kernel, which must necessarily be completed before calling a cuBLAS\textsuperscript{\textregistered} function, at the end of which the processed data should be transferred back to the CPU. All it takes is assigning the sequentially dependent GPU function calls to the same stream. Multiple streams can be created and used at the same time, the obtained behaviour mimics batch job submissions to supercomputing facilities. CUDA\textsuperscript{\textregistered} guarantees the correct execution line and synchronization within the same stream, whereas different streams must instead be synchronized with each other by hand. The compiler typically schedules executions and memory transactions so that the GPU use is maximized, superposing different GPU function calls and data transfers from separate streams \cite{CUDA}.

Despite the lower latency, optimized programs are designed to also control the way threads access shared memory locations. If two adjacent threads are asked to access two non-contiguous array elements, then the memory access is performed on two cycles instead of a single one, slightly slowing down the program execution. This issue is called \textit{bank conflict}, and can be avoided ensuring thread-varying elements to be stored in the leading dimension of shared memory arrays. Bank conflicts do not happen if all threads access the same single memory location.

\section{Implementation}\label{sec:implementation}
Among the implementations outlined in this section, the case of independent PC runs for all the propagated trajectories is considered as benchmark. This allows to directly assess the performance of the augmented PC algorithm against the original integrator for the same test case. All the algorithms were implemented using the C language, with the sole exception of the GPU program that was coded in CUDA.

\subsection{Independent runs PC workflow}
\begin{figure}[h!]
	\tiny
	\centering 
	\def\svgwidth{\columnwidth}
	\includegraphics[width=\columnwidth]{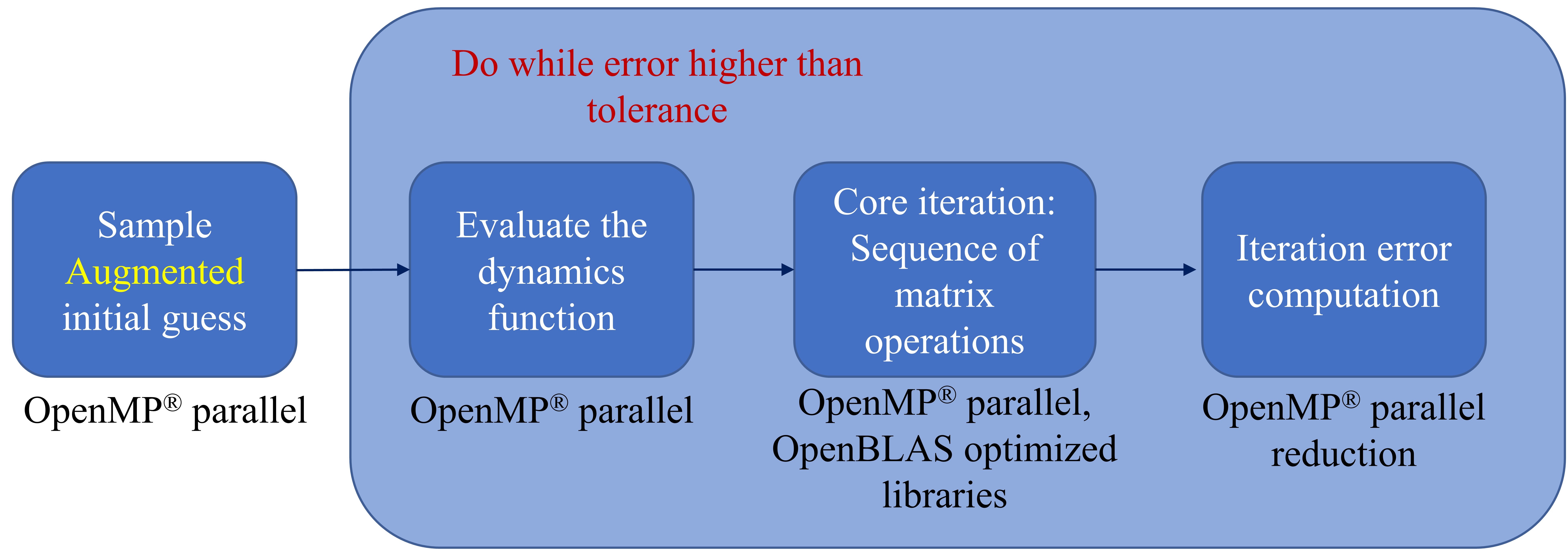}
	\caption{Standard PC workflow.}
	\label{fig:pcbasic}
\end{figure}

The basic workflow of the independent PC runs is given in the block-scheme of Figure \ref{fig:pcbasic}. The only parallelization possibilities, for high numbers of trajectories, apply at the highest level, inevitably introducing a considerable overhead for both the inner sequential execution and the still parallelizable inner functions. In fact, all the per-trajectory steps of the PC process would still be parallelizable algorithms per se.

\subsection{Sequential Augmented PC workflow}
\begin{figure}[h!]
	\tiny
	\centering 
	\def\svgwidth{\columnwidth}
	\includegraphics[width=\columnwidth]{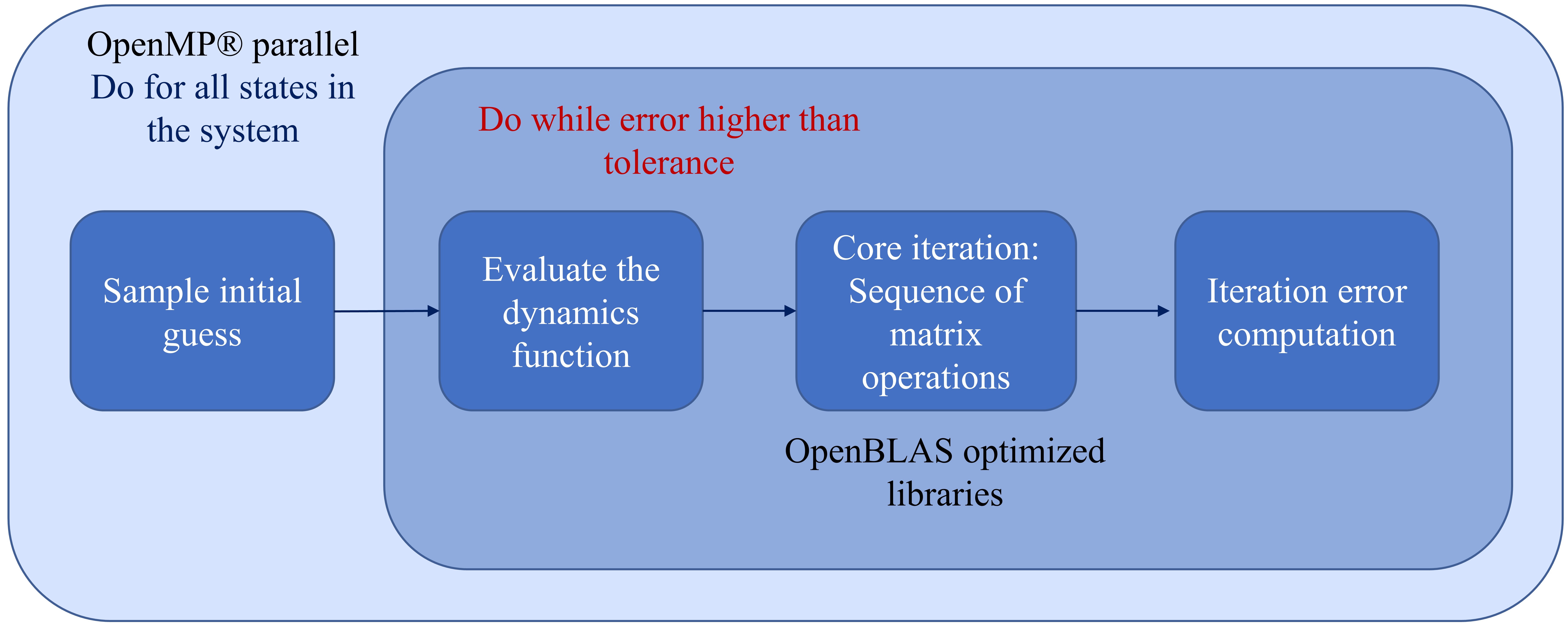}
	\caption{Augmented PC workflow.}
	\label{fig:pcaug}
\end{figure}

The implementation of the augmented PC integration follows a one-level augmentation only, to highlight the pipeline benefits in terms of overhead that this implementation introduces. A block-scheme representation of the augmented PC integration workflow is given in Figure \ref{fig:pcaug}. The conceptual change, from the PC iteration viewpoint, is only the initial sampling in a single array containing all the state vectors of all the trajectories of the augmented system.

\subsection{OpenMP\textsuperscript{\textregistered} parallelized Augmented PC workflow}
\begin{figure}[h!]
	\tiny
	\centering 
	\def\svgwidth{\columnwidth}
	\includegraphics[width=\columnwidth]{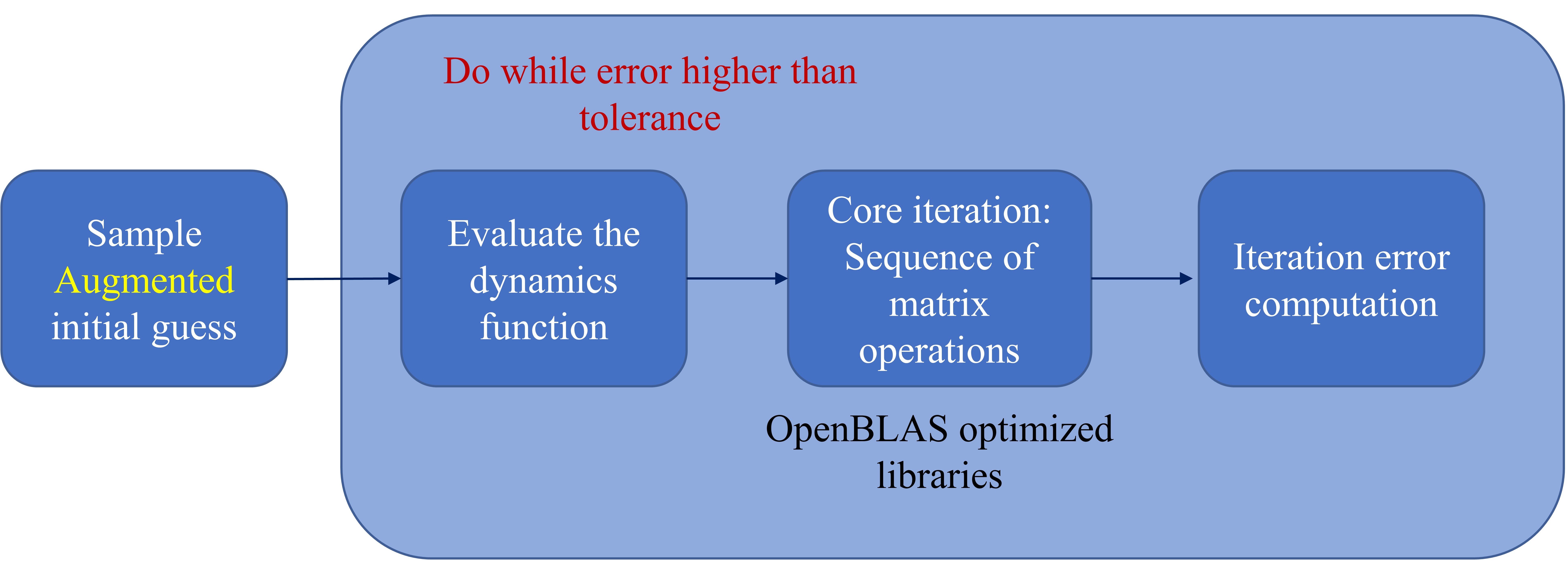}
	\caption{Augmented and OpenMP\textsuperscript{\textregistered} parallelized PC workflow.}
	\label{fig:pcaugparallel}
\end{figure}

The parallelization of the augmented system integration, whose block-scheme representation is given in Figure \ref{fig:pcaugparallel}, becomes fine-grained. It acts directly on the single state vectors for the dynamics function evaluation and on the elementary products of the matrix operations, moreover already implemented by the OpenBLAS libraries \cite{openblas}. In addition, reduction operations can be made through OpenMP\textsuperscript{\textregistered} for a cooperated and parallel search of the maximum error.

\subsection{CUDA\textsuperscript{\textregistered} Augmented PC workflow}
\begin{figure}[h!]
	\tiny
	\centering 
	\def\svgwidth{\columnwidth}
	\includegraphics[width=\columnwidth]{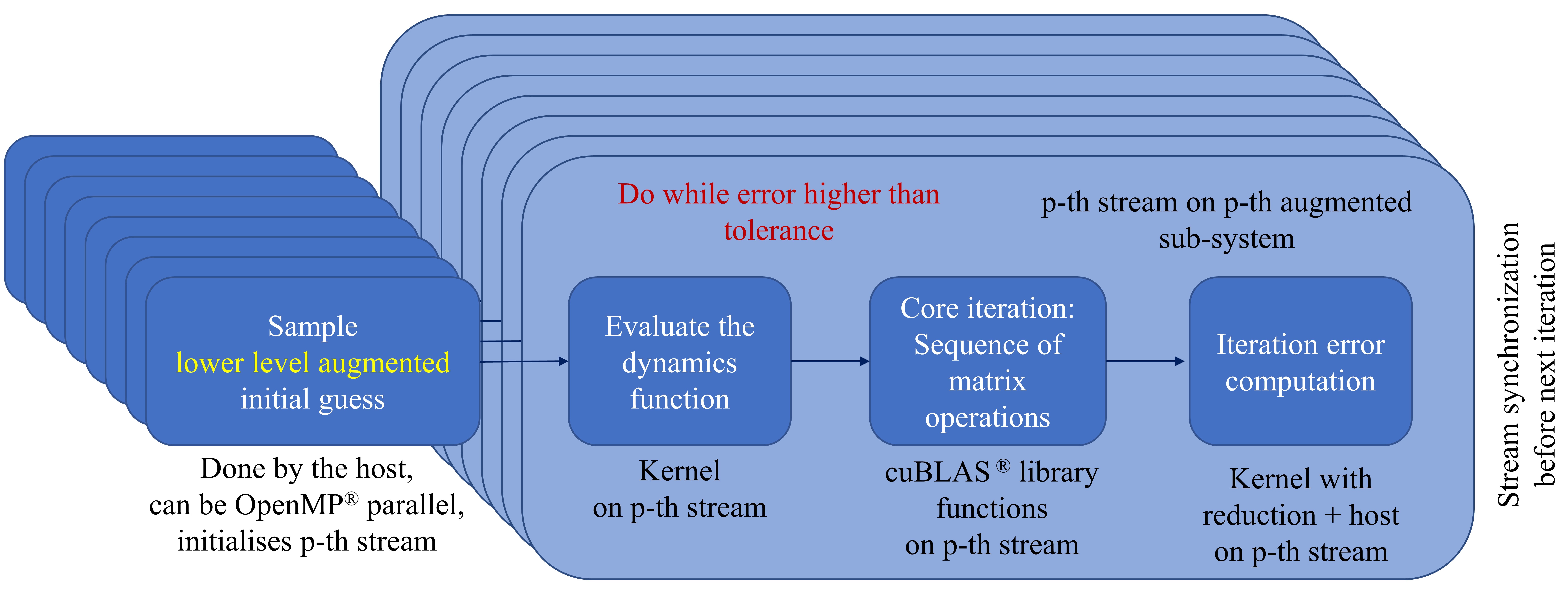}
	\caption{Augmented CUDA\textsuperscript{\textregistered} PC workflow.}
	\label{fig:pcaugcuda}
\end{figure}

The block-scheme representation of the CUDA\textsuperscript{\textregistered} algorithm is given in Figure \ref{fig:pcaugcuda}. The two-level augmentation concept is exploited, assigning one higher level augmented system to each CUDA\textsuperscript{\textregistered} stream and using the thread-based parallelism on the lower level augmented systems.

The principal benefit is the cooperation between CPU and GPU for the overall execution, with as many operations as possible executed concurrently. Each lower level augmented system is initially sampled by the host and then moved to the GPU. The CUDA\textsuperscript{\textregistered} stream management API allows to overlap the CPU sampling of the next higher level augmented systems with memory transfers and kernel executions of the already launched ones\footnote{This is true as long as the CPU memory is allocated as \textit{paged} with specific CUDA\textsuperscript{\textregistered} functions \cite{CUDA}.}. Similarly, the last step of the PC iteration requires to retrieve the computed iteration error for each stream from the GPU to the CPU for loop control purposes, which is also subject to the stream concurrency benefits. A stream synchronization at the end of each while loop iteration is necessary to achieve the overlapping behaviour of all the streams. Running independent loops for each higher level augmented system would result in completely sequential and non-overlapped executions. If a single CUDA\textsuperscript{\textregistered} stream is generated, the standard one-level augmented system case is reproduced, albeit with the GPU computing acceleration instead of the OpenMP\textsuperscript{\textregistered} implementation previously described.

The warp-centric programming model of CUDA\textsuperscript{\textregistered} kernels requires a small modification on the lower level augmented system definition. Contiguous array elements should be of the same component type (i.e. contiguous $x$ coordinates, then contiguous $y$ coordinates, and so on), instead of storing state vector by state vector. This aspect might seem an implementation detail, however it is fundamental to ensure coalesced global memory access. A too high latency would happen otherwise, which cannot be hidden even by intensive parallelized GPU computations. The just discussed modification has no effect on the overall algorithm structure, all it requires is the dynamics and error kernels to be implemented following this array element logic. This aspect is discussed in more detail in the following section, together with the implications it has especially on the evaluation of the dynamics function.

\subsubsection{Dynamics model, array sorting, and CUDA kernel}
At the core of the PC integration scheme lies the evaluation of the dynamics function at each Picard iteration. This task can be performed in parallel for all the states of the system being integrated, however, although conceptually simple, its implementation may not be straightforward in the GPU computing case. The more complex the dynamical model becomes, the more intertwined its implementation inevitably gets, possibly  requiring to access data distributed in multiple arrays, possibly of notably different sizes. The accuracy requirements of the proposed application demand to work under the restricted relativistic N-body problem, following the Einstein-Infeld-Hoffmann equations \cite{Seidelmann1992}, which has a dynamics function of the form:
\begin{equation}
	\begin{gathered}
		\ddot{\textbf{r}} = \textbf{f}\big(\textbf{r},\dot{\textbf{r}},\textbf{r}_i,\dot{\textbf{r}}_i,\ddot{\textbf{r}}_i\big)
	\end{gathered}
\end{equation}
with $i$ condensing the dependence on the states of all the major bodies in the ephemeris model, e.g. the solar system planets, and $\textbf{r}$, $\dot{\textbf{r}}$, $\ddot{\textbf{r}}$ denoting position, velocity, and acceleration, respectively. The relations are unfortunately non-linear: as a consequence, each CUDA\textsuperscript{\textregistered} thread cannot perform simple operations on one single array element, since both position and velocity of each state are required to compute any acceleration component. Moreover, ephemerides data for $\textbf{r}_i$, $\dot{\textbf{r}}_i$, $\ddot{\textbf{r}}_i$ also enter the dynamics function. These aspects suggest to implement the dynamics kernel having each CUDA\textsuperscript{\textregistered} thread to process one full state vector, rather than one element. At the same time, coalescing the global memory access remains crucial to obtain a well-performing kernel. The PC method introduces however a partial constraint on the array shapes: the matrix multiplications that build the method need the sampled trajectory states to be stored as rows of an overall matrix, fixing the different times to identify each row. For a column-major sorted augmented state matrix, contiguous state elements are interrupted by the ending time nodes, likely leading to non-coalesced memory access for numerous warps. Row-major sorted state arrays feature instead non-coalesced access for all the state elements.

To cope with these issues, the lower-level augmented state is reformulated by stacking along the columns the same components of all the state vectors in the augmented system: 
\begin{equation}
	\begin{gathered}
		\textbf{Y}^{(i)}_{j} = \bigg[x^{(i)}_{j,1} \enskip \cdots \enskip x^{(i)}_{j,M} \enskip \cdots \enskip p^{(i)}_{j,1} \enskip \cdots \enskip p^{(i)}_{j,M} \enskip \cdots \enskip \dot{z}^{(i)}_{j,1} \enskip \cdots \enskip \dot{z}^{(i)}_{j,M} \bigg], \\ j=1,...,N, \enskip p=y,z,\dot{x},\dot{y}
	\end{gathered}
\end{equation}
where $(x,y,z)$ are the Cartesian components of $\textbf{r}$. The advantage is obvious in the column-major sorting case, since all the common components are found in adjacent memory addresses. Since in the presented application the number of states in the augmented system is much larger than the number of sampled trajectory nodes, many contiguous state components also appear in the row-major sorted array case. In addition, the higher-level augmented system definition may remain unaltered, since different CUDA\textsuperscript{\textregistered} streams would be processing each lower-level sub-systems. The augmented force matrix $\textbf{F}^{(i)}_j$ can be adapted accordingly, without introducing any modification to the matrix multiplication characterizing the PC iteration. Lastly, bank conflicts (explained in Section \ref{sec:CUDAadvanced}) are automatically avoided \cite{CUDA} with this array sorting approach.

The kernel design is tied with the array sorting strategy. In particular, a key role is played by the fixed time nodes. The augmented system logic is a consequence of the shared time nodes among the different state vectors, this feature should also be exploited to design the thread blocks to make the most of the available shared memory. In particular, the amount of memory required to store the planetary ephemerides is minimized if all the threads in a block process state vectors corresponding to the same time node. For this reason, the proposed program implements a row-major sorting strategy of state and force matrices. The augmented state matrix is accessed as a two-dimensional block array: one dimension (the rows) follows the different time nodes, whereas the other is used to split the states of a common time node into smaller chunks, each containing 32 states\footnote{32 is the warp size for most NVIDIA\textsuperscript{\textregistered} graphics cards \cite{CUDA}. For augmented systems with a number of states that is not a multiple of the block size, the dynamics kernel is implemented so that the last block of threads processes the remainder of the integer division between the number of states and the block size.}. In this way, all the states in the same block of the two-dimensional block array require exactly the same ephemerides data, because they are all associated to the same time node. For fixed-time thread blocks bank conflict is automatically avoided also when reading ephemerides data from the shared memory, since all the threads are forced to access the same ephemerides item or vector component \cite{CUDA}.

\begin{figure}[h!]
	\tiny
	\centering 
	\def\svgwidth{\columnwidth}
	\includegraphics[scale=0.55]{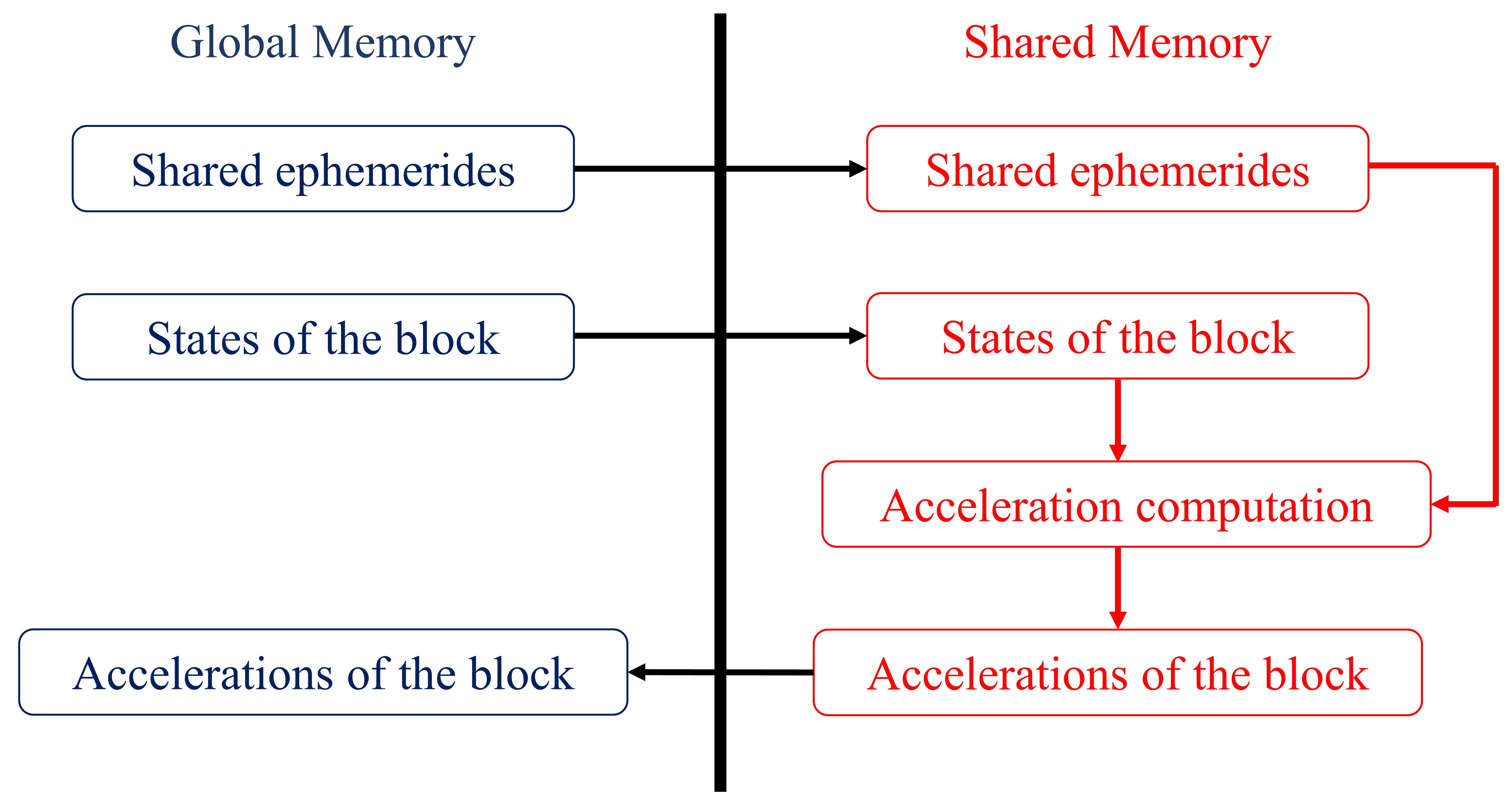}
	\caption{Dynamics kernel memory management.}
	\label{fig:dynkernel}
\end{figure}

The computation of the dynamics function is a compute-bound task: most of the effort lies on performing computations on a limited amount of data, rather than on the movement of a large amount of information between two memory locations. Furthermore, the values of the state elements need to be repeatedly accessed. For these reasons and because of the limited capacities of thread-private registers, shared memory is used to also temporarily store the state values, because of its lower latency compared to the global GPU memory \cite{CUDA}. The resulting dynamics kernel is summarized in Figure \ref{fig:dynkernel}.

In addition, a close look at the Einstein-Infeld-Hoffmann equations reveals that the acceleration of each propagated body depends not only on its state and the gravitational parameter and states of major bodies in the ephemeris model, but also explicitly on the gravitational potential and the acceleration of such bodies \cite{Seidelmann1992}. These contributions can be computed before the evaluation of the dynamics function itself in the restricted problem of the proposed application, saving the computational burden of a task that would be repeated at each Picard iteration. The values of gravitational potential and the acceleration of the major bodies of the ephemeris model are computed by the CPU before starting the Picard iterations, then moved to the global GPU memory, and eventually loaded to the shared memory for the time node associated with each block, along with the ephemeris states and gravitational parameters.

\subsubsection{CPU-GPU cooperated iteration error computation}\label{subsec:error}
All the involved CUDA\textsuperscript{\textregistered} kernels are run at each Picard iteration. Their execution must be called by the CPU, which also stops the Picard iteration while loop as the error between two consecutive steps falls below the desired tolerance. Since the updated states already reside on the GPU, the GPU massive parallelism can be exploited to accelerate the computation of the iteration error, transferring only a limited amount of data to the CPU to be used for the loop control. Despite dealing with the augmented state system, the error is still defined on a per-state basis, with the maximum of the errors of all the states that is used to control the loop.

The error computation process involves two separate kernels and a CPU function. The first kernel computes both the position and the velocity errors and stores the maximum between these two, for all the states in the augmented system. The second kernel computes the maximum error of groups of 4096 states: 1024 thread-sized blocks are created, discerning the first maximums while reading four consecutive chunks of state errors into the shared memory. Then, 1024 threads cooperate to find the actual maximum error among the remaining 1024 state errors, with reduction-driven parallelism\footnote{Like the dynamics kernel, the error kernels are implemented so that the last thread block processes the remainder between the integer division between the number of states and the block size (4096 in the case of the reduction kernel). The reduction steps are controlled accordingly: only those corresponding to a number of threads less than or equal to the number of states in the block are activated.}. Eventually, the maximum error is copied back to the global memory, in a new array consisting of reduced errors only. Finally, this whole array is copied back to the CPU, which finds the actual maximum with a traditional sequential for loop-based approach. Even with augmented states made of millions of state vectors, this approach makes the CPU search sequentially only over hundreds to thousands candidates at most, with a negligible computational cost compared to the other steps.

\section{Program performance}\label{sec:performance}

\subsection{Test case}\label{sec:testcase}
\begin{figure}[h!]
	\tiny
	\centering 
	\def\svgwidth{\columnwidth}
	\includegraphics[scale=0.65]{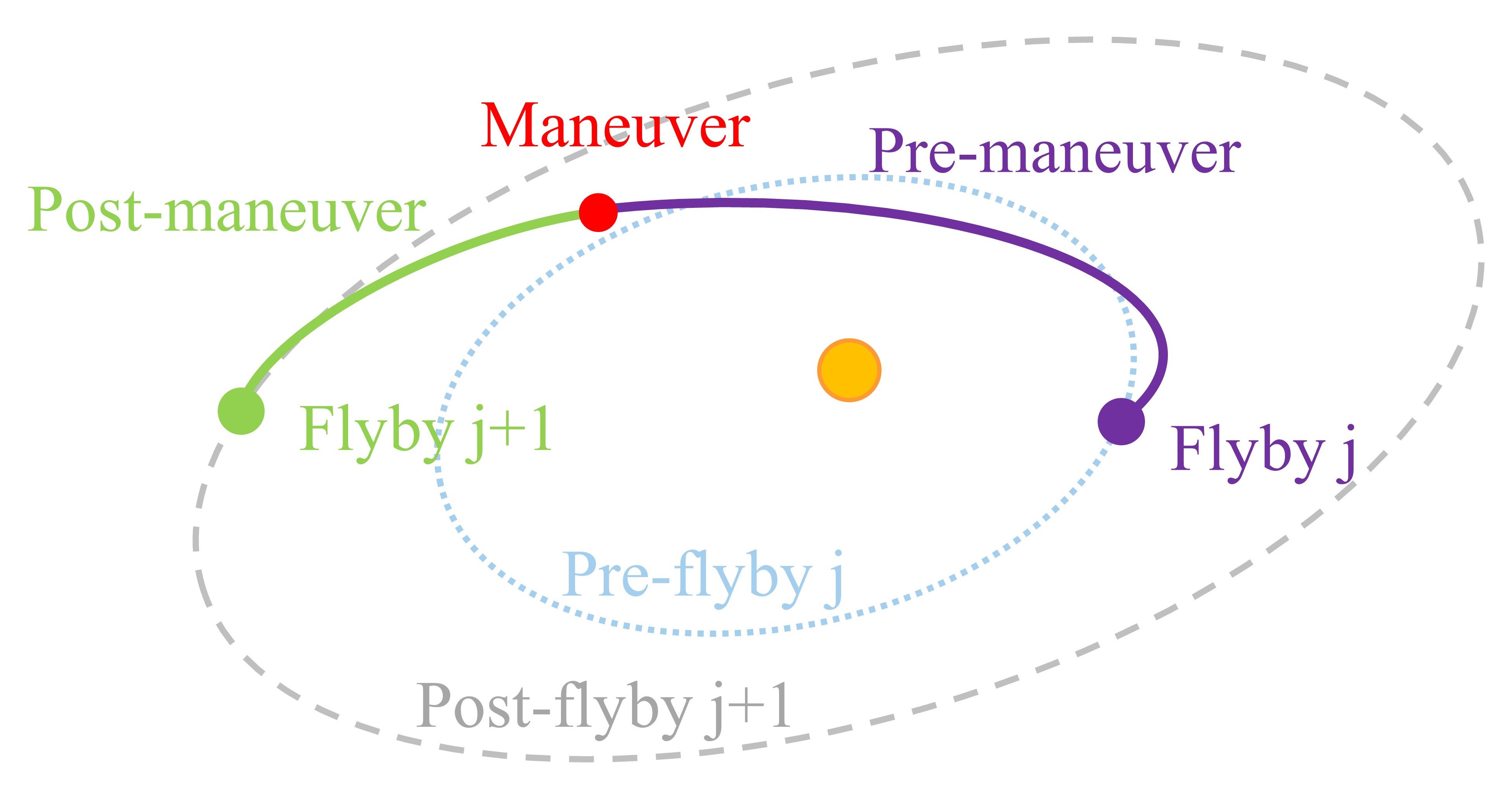}
	\caption{Orbital scheme of the phase from flyby $j$ to flyby $j+1$.}
	\label{fig:legscheme}
\end{figure}

The test case follows what already computed in \cite{Masat2021AASAIAA,Masat2022acta}. The first resonant phase with Venus of a Solar Orbiter-like mission was reproduced, designing a continuous trajectory even during flyby injections and exits. The approach is based on the b-plane \cite{Valsecchi2003} description of flybys, and uses the PC method for efficient fixed-point simulations to surf the relativistic N-body environment\footnote{The enforced trajectory continuity even during flybys requires extreme design precision, because the fast dynamics of close approaches would lead to divergent and therefore meaningless trajectories, already after the first flyby.} minimizing the deep space maneuver effort required to control the flyby entrance. Following the trajectory scheme of Figure \ref{fig:legscheme}, the design algorithm follows a backward, dynamic programming-like recursion logic, i.e. designing flyby $j+1$ before flyby $j$, to preserve accuracy and continuity while also fulfilling mission requirements. The reader can find more insight on the design algorithm and technique in \cite{Masat2021AASAIAA,Masat2022acta}.

\begin{figure}[h!]
  \centering
  \subfloat[Deep space correction maneuver and overview.]{\includegraphics[width=0.49\textwidth,trim= 12cm 0cm 12cm 0cm, clip]{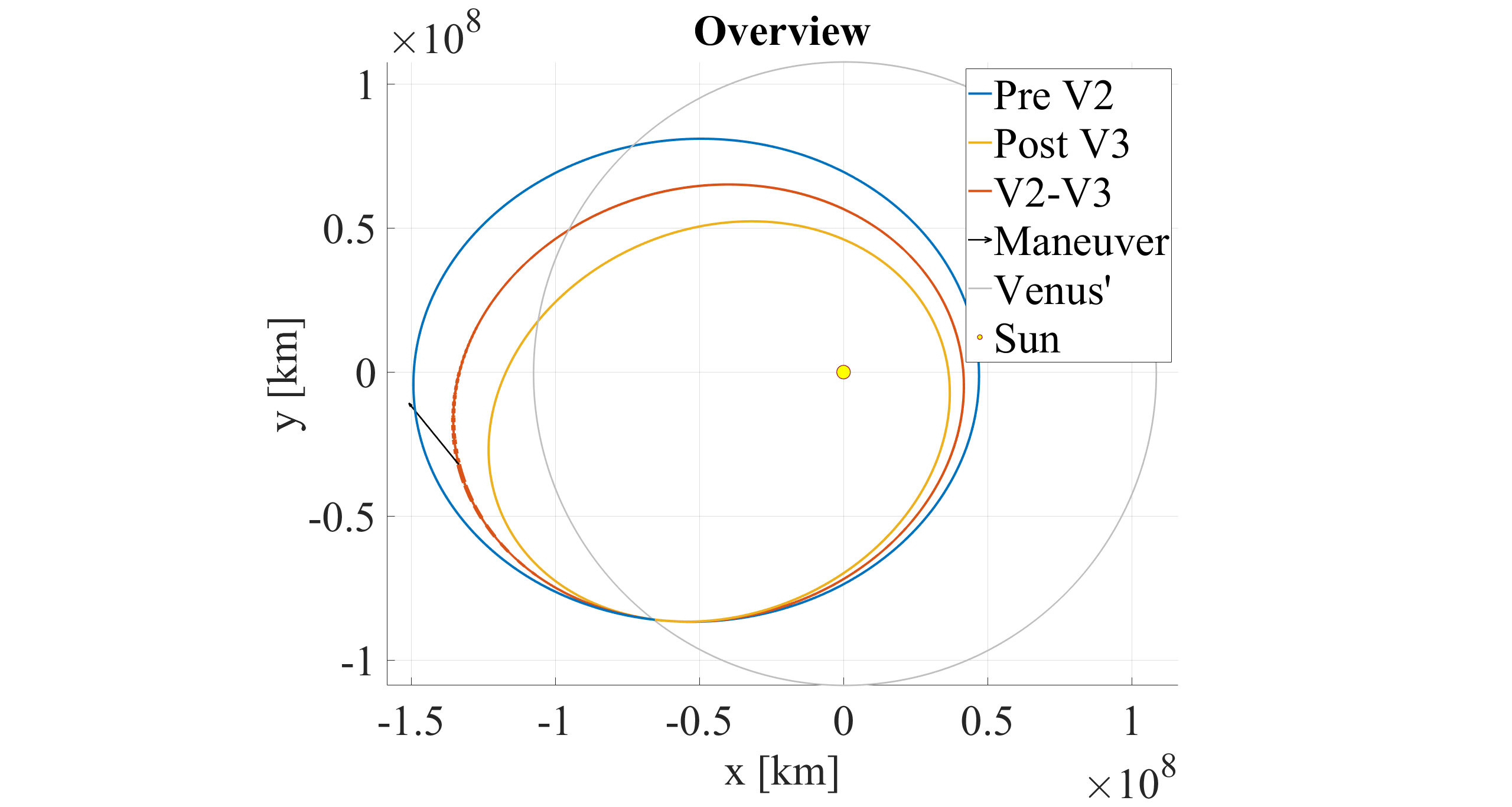}\label{fig:contoverview}}
  \subfloat[Zoom over Venus' flybys V2 and V3.]{\includegraphics[width=0.48\textwidth,trim= 12cm 0cm 14cm 0cm, clip]{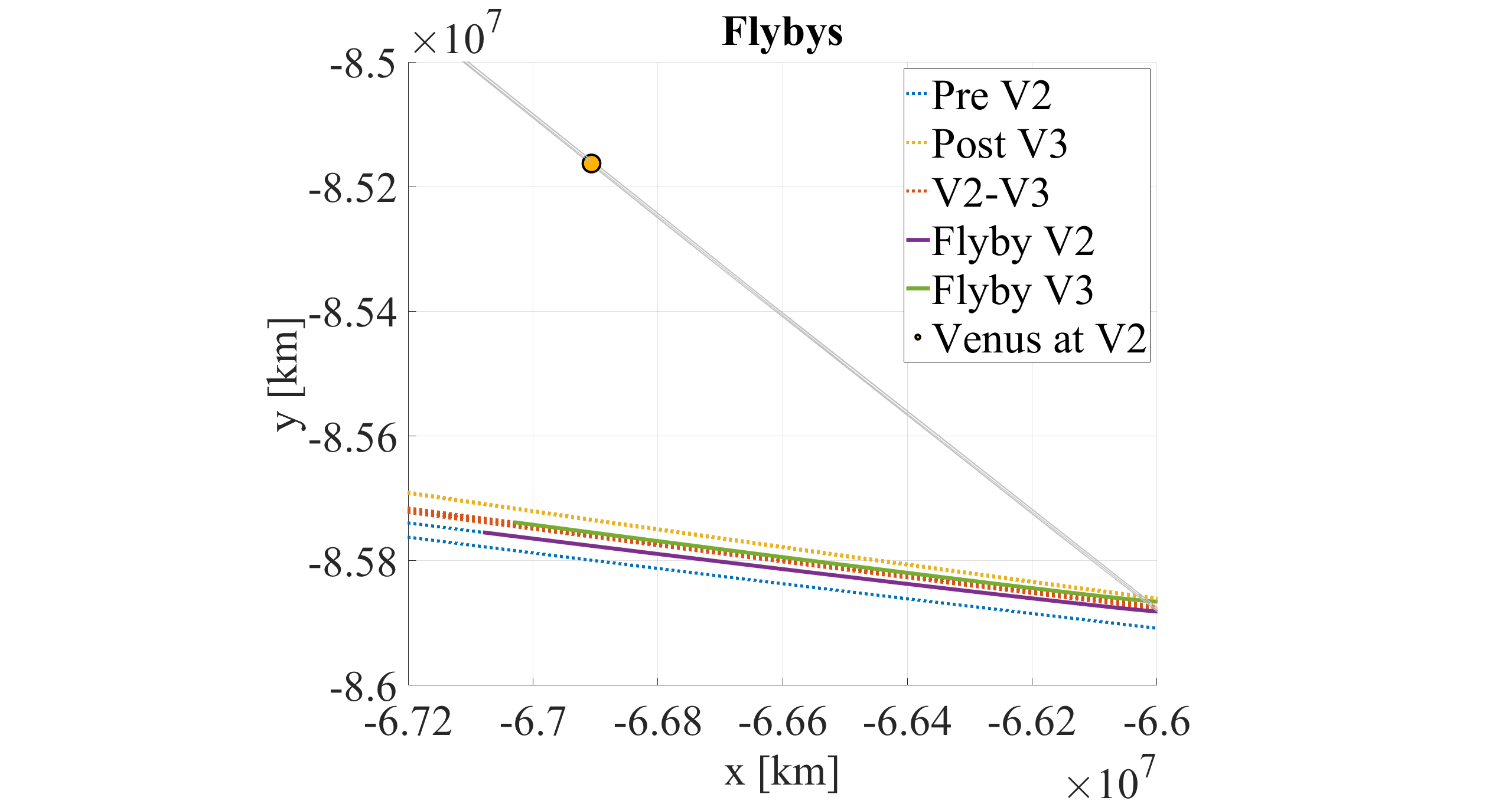}\label{fig:contfly}}
  \caption{Solar Orbiter's continuous first resonant phase with Venus. Pictures from \cite{Masat2021AASAIAA,Masat2022acta}.}
\end{figure}

The Matlab\textsuperscript{\textregistered}\footnote{Following the update to Matlab\textsuperscript{\textregistered} R2021b higher runtimes were obtained for the same final results, although not affecting the presented discussion since the performance analysis is only made on the C and the CUDA algorithm versions.} implementation proposed in \cite{Masat2021AASAIAA,Masat2022acta} was re-run on a single core of a local workstation equipped with an Intel\textsuperscript{\textregistered} Core\textsuperscript{TM} i7-7700 CPU (3.60 GHz), running Windows\textsuperscript{\textregistered} 10 Pro. The algorithm converged running $13509$ PC integrations in $506.3$ seconds, to the residual $\Delta\textbf{r}^*$ and impulsive action $\Delta\textbf{v}^*$ for the required maneuver presented in Table \ref{tab:resultsdrdv}.
\begin{table}[h!]
	\centering 
	\def\svgwidth{\columnwidth}
	\caption{Optimization results, in terms of position difference residual $\Delta\textbf{r}^*$ and correction effort $\Delta\textbf{v}^*$ at the maneuvering time corresponding to the apocenter of the post-flyby orbit.}
	\begin{tabular}{c c c c c c}
		\multicolumn{3}{c}{$\Delta\textbf{r}^*$} & \multicolumn{3}{c}{$\Delta\textbf{v}^*$} \\
		\multicolumn{3}{c}{$[\text{m}]$} & \multicolumn{3}{c}{$[\text{m/s}]$} \\
		\hline		
		$x$ & $y$ & $z$ & $x$ & $y$ & $z$ \\
		\hline
		$-0.52$ & $-0.52$ & $-1.19$ & $-1.28$ & $1.57$ & $0.22$ \\
		\hline
	\end{tabular}
	\label{tab:resultsdrdv}
\end{table}
To better detail the embedded PC integrations, they all feature $200$ Chebyshev nodes spanning the arc connecting the flyby exit and the maneuver point, long about 0.87 orbital periods. Figures \ref{fig:contoverview} and \ref{fig:contfly} show the final trajectory, optimized in the relativistic N-body environment.

\subsection{Computational setup}
To build a common framework for the pure algorithm performance evaluations, the 13509 initial conditions generated in the optimization process to eventually obtain the results of Table \ref{tab:resultsdrdv} are re-run using C and CUDA implementations of the PC integration. The execution of the 13509 independent runs with the C implementation of the algorithm is considered as benchmark case, both completely sequential and parallelized with OpenMP \cite{openmp}. The matrix operations featured in the PC iterations are performed using the OpenBLAS library \cite{openblas,openblas1,openblas2}. All the presented runs of the C algorithm have been executed on a machine running Ubuntu Linux 20.04 equipped with 40 physical / 80 logical cores of the type Intel\textsuperscript{\textregistered} Xeon\textsuperscript{\texttrademark} CPU E5-4620 V4 running at 2.1 GHz, with varying number of OpenMP threads and the "\texttt{o3}" \texttt{gcc} compiler optimization enabled. Because of the physical machine where the GPU was available, the CUDA\textsuperscript{\texttrademark} code has been run on the same workstation of the Matlab\textsuperscript{\textregistered} optimal solution computation. In this case a four core OpenMP\textsuperscript{\textregistered} parallelization on a Intel\textsuperscript{\textregistered} Core\textsuperscript{TM} i7-7700 CPU (3.60 GHz), is combined for concurrent executions with a NVIDIA\textsuperscript{\textregistered} GTX 1050 (1.3GHz) graphics card\footnote{This is a 2016 low-end gaming card model, whose design purpose is far from the double precision computing of this work. Modern gaming/professional cards could run up to 60 times faster, data center cards up to 400-500 times faster than this model for the presented application.}. As a final remark, in all the presented cases the relative error among the different implementations for corresponding initial conditions always falls below the specified PC relative tolerance ($10^{-12}$), allowing runtime-only performance comparisons. These small differences are due to the inevitably slightly different order of the elementary operations performed by the different machines, libraries and implementations.

\subsection{Accuracy comparison}
\begin{figure}[h!]
	\tiny
	\centering 
	\def\svgwidth{\columnwidth}
	\includegraphics[width=\columnwidth]{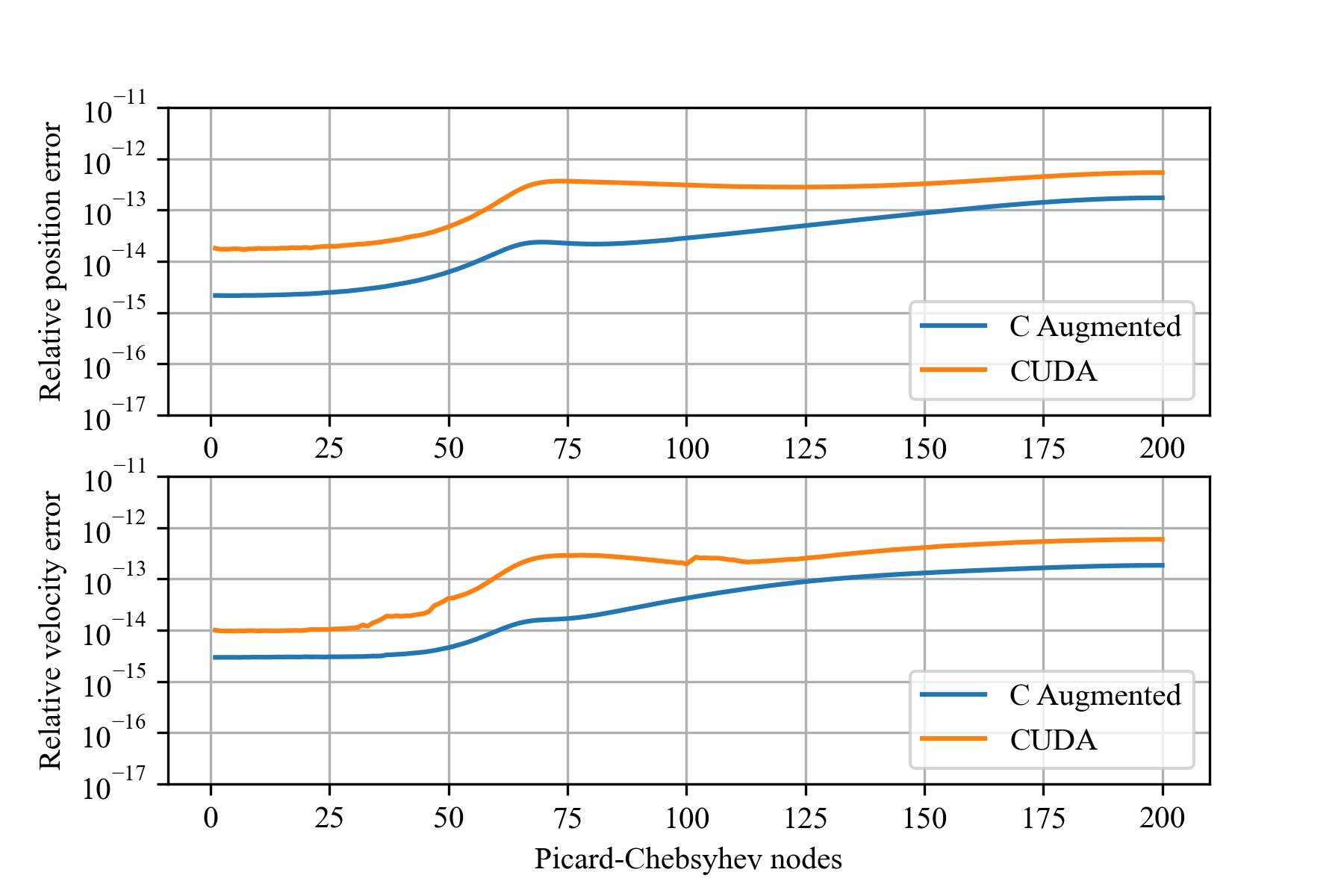}
	\caption{Errors of C augmented and CUDA\textsuperscript{\textregistered} programs, as the average of the error of all the states in the augmented systems.}
	\label{fig:properror}
\end{figure}

Figure \ref{fig:properror} shows the evolution of the propagation error, measured as relative position and velocity error with respect to the independent runs case, for both the C augmented and the CUDA\textsuperscript{\textregistered} programs. In both cases the error remains lower than the specified relative tolerance used to halt the Picard-Chebyshev iterations, set to $10^{-12}$.

\begin{figure}[h!]
	\tiny
	\centering 
	\def\svgwidth{\columnwidth}
	\includegraphics[width=\columnwidth]{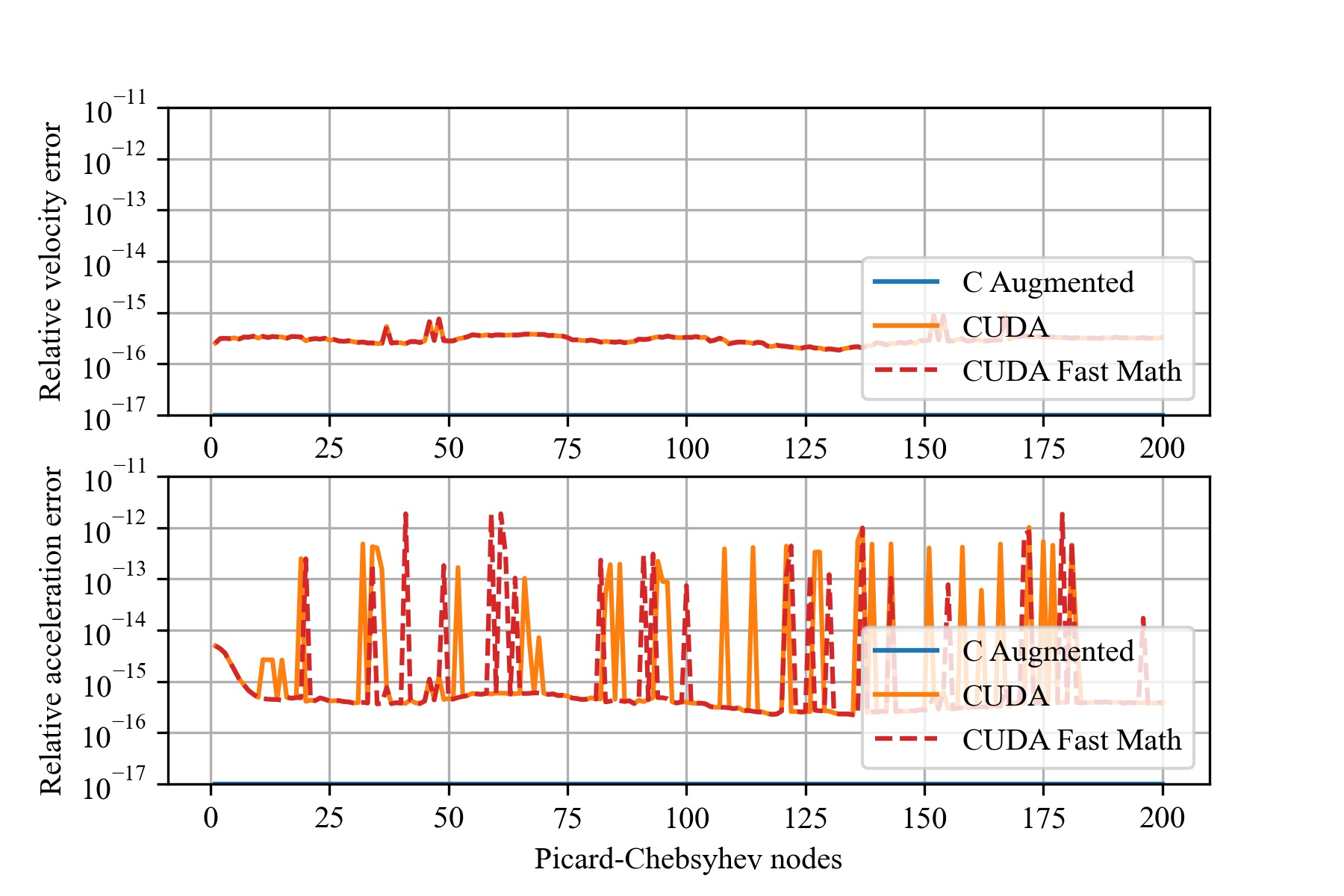}
	\caption{Errors of C augmented and CUDA\textsuperscript{\textregistered} programs, as the average of the error of all the states in the augmented systems.}
	\label{fig:dynerror}
\end{figure}

Figure \ref{fig:dynerror} shows instead the evolution of the error for the dynamics function only, distinguishing two different compilations of the CUDA\textsuperscript{\textregistered} programs, with (red dashed) and without (orange) the \texttt{--use-fast-math} flag enabled. Despite the error experienced in Figure \ref{fig:properror}, the C augmented program shows no error for the computation of the dynamics function. That means, the error is accumulated throughout the Picard-Chebyshev iterations only because of the different structure of the matrix multiplications. Their implementation in the OpenBLAS library \cite{openblas,openblas1,openblas2} is optimized on the matrix size. This is a known issue in high-performance computing problems, where the floating point representation of numbers interferes with a change in the order of the basic math operations on the matrix elements, producing small deviations with respect to a reference non-parallelized solutions. No differences were observed between parallelizing and not parallelizing the augmented C programs. On top of this consideration, the CUDA\textsuperscript{\textregistered} program is also subject to the errors introduced by the different compilers. Other than being two different environments (Ubuntu Linux with GNU\textsuperscript{\textregistered} compilers for the C augmented program, Windows 10 Pro with Microsoft\textsuperscript{\textregistered} Visual Studio\textsuperscript{\textregistered} and Nvidia\textsuperscript{\textregistered} CUDA\textsuperscript{\textregistered} compilers for the CUDA \textsuperscript{\textregistered} program), small differences can also be observed by setting different optimization flags in the compilation. The comparison between enabling and disabling the fast math optimization options are run to exclude the possibility of implementation problems for the dynamics kernel. As it can be observed comparing the two compilations, the error spikes reaching $10^{-12}$ happen in a seemingly unpredictable manner and can be completely attributed to the compiler, because happening at different Picard-Chebyshev nodes for the two cases. Otherwise, the error level remains more than two orders of magnitude lower. In any case, the accumulated effect of this error source is not taking the overall error above the $10^{-12}$ iteration tolerance, as Figure \ref{fig:properror} already highlighted.

\subsection{Performance comparison}
Table \ref{tab:seqruntime} shows the runtime difference between the sequential C programs. The improved efficiency of the augmented integration can be immediately seen even in this sequential case, where the augmented system approach runs $23.87\%$ faster, because of the minimized overhead experienced by sharing the outer while loop. In the considered application the various trajectories all require 41 or 42 PC iterations, making it negligible to keep running trajectories even if their PC process has already converged.
\begin{table}[h!]
	\centering 
	\def\svgwidth{\columnwidth}
	\caption{Sequential runtimes for the independent runs and the augmented system executions.}
	\begin{tabular}{c c}
		\textsc{\textbf{Case}} & \textsc{\textbf{Runtime}} \\
		& [s] \\
		\hline
		Independent runs & 245.02 \\
		\hline
		Augmented system & 186.54 \\
		\hline
	\end{tabular}
	\label{tab:seqruntime}
\end{table}

\begin{figure}[h!]
	\tiny
	\centering 
	\def\svgwidth{\columnwidth}
	\includegraphics[width=\columnwidth]{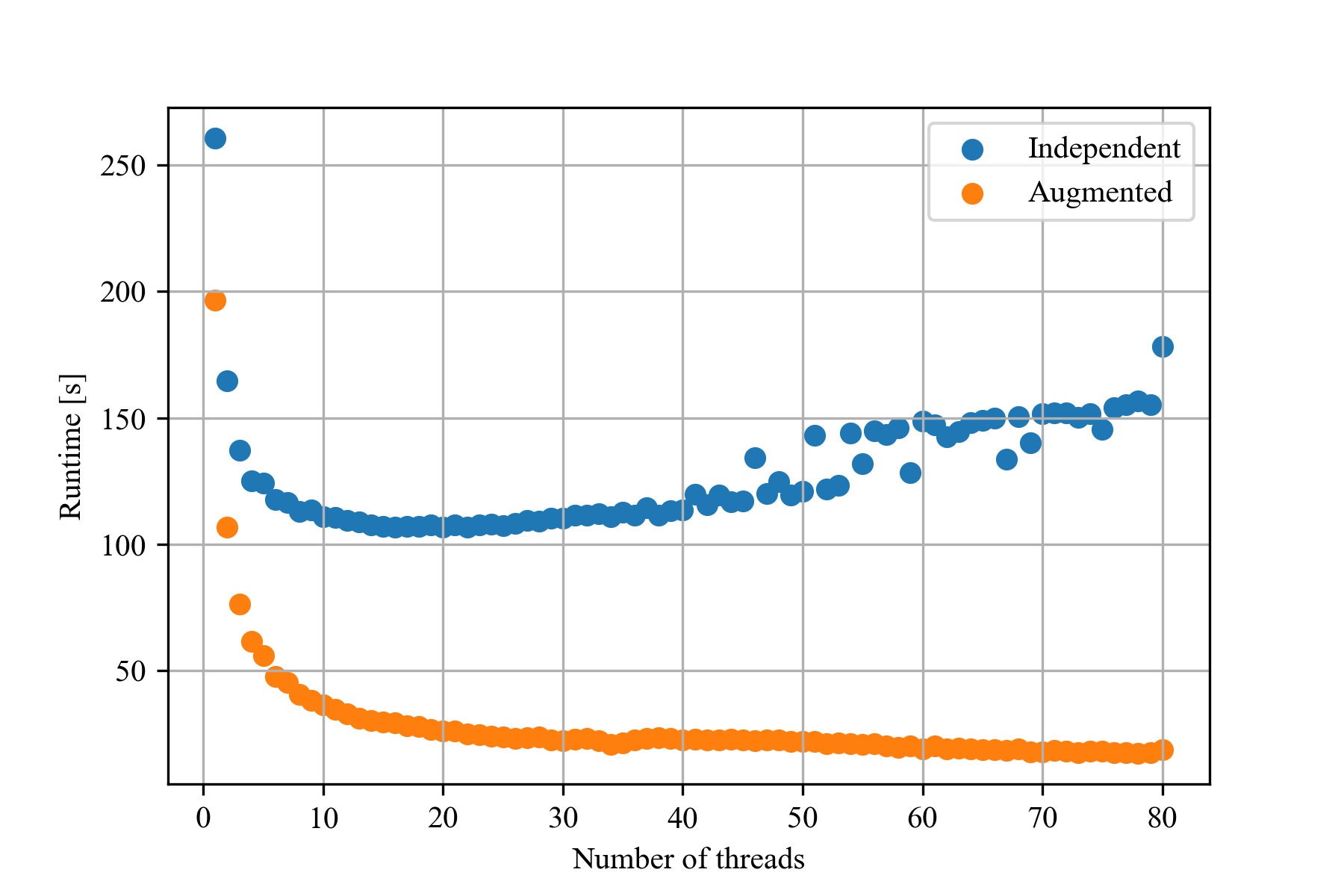}
	\caption{Augmented system and independent integrations C code runtime comparison with OpenMP\textsuperscript{\textregistered} parallelization and varying number of threads.}
	\label{fig:runtimecompare}
\end{figure}

The scalability properties of the integration of independent trajectories and the augmented system are studied on the C implementations, i.e. assessing how well the execution of the two programs accelerates with increasing number of OpenMP\textsuperscript{\textregistered} threads. Figure \ref{fig:runtimecompare} shows that the augmented system features excellent scalability properties, for a runtime that keeps decreasing for increasing number of OpenMP\textsuperscript{\textregistered} threads. On the contrary, the integration of independent trajectories experiences even higher runtimes after a certain number of threads. This happens because the parallelization itself introduces some overhead to the overall program execution, which cannot be compensated for by newly created parallel threads. 

\begin{figure}[h!]
	\tiny
	\centering 
	\def\svgwidth{\columnwidth}
	\includegraphics[width=\columnwidth]{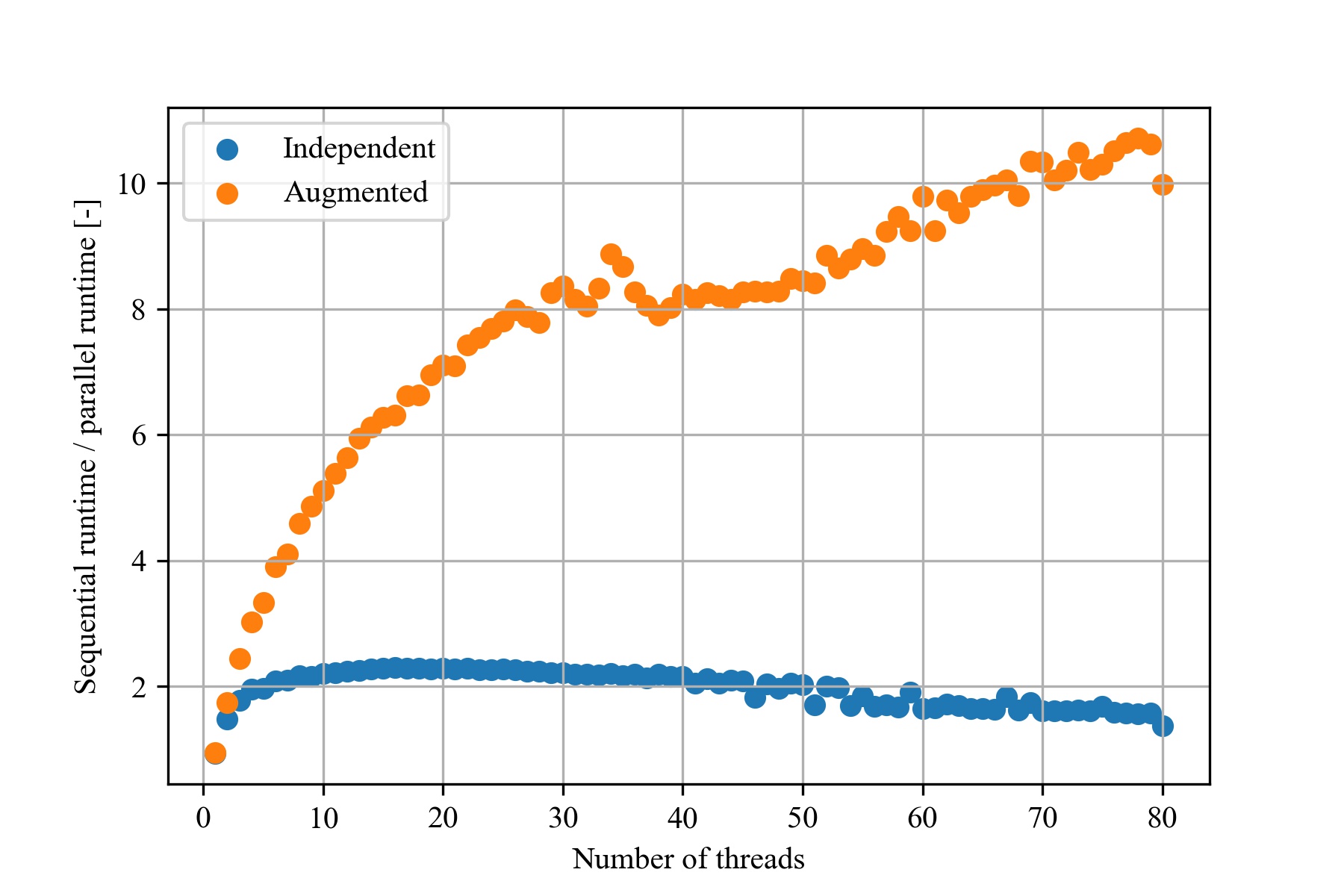}
	\caption{Augmented system and independent integrations C code speedup comparison with OpenMP\textsuperscript{\textregistered} parallelization and varying number of threads.}
	\label{fig:speedupcompare}
\end{figure}

Figure \ref{fig:speedupcompare} shows the achieved speedup, defined as the ratio between the sequential runtime and the parallel runtime, varying the number of OpenMP\textsuperscript{\textregistered} threads. It provides a measure of how parallelizable the algorithm is, with higher values underlining higher accelerations. The augmented system integration shows once again excellent scalability properties even at high number of threads, suggesting the efficiency of the GPU computing transition even without having assessed the performances of the CUDA\textsuperscript{\textregistered} implementation yet.
If the number of threads is set equal to one, the algorithm conceptually reduces to a sequential case. However, enabling compiler optimization and the OpenMP\textsuperscript{\textregistered} flag in the compilation, a parallel program introduces data distribution and retrieval tasks across the possibly multiple workers. With OpenMP\textsuperscript{\textregistered}, the number of threads is specified by an environment variable right before running the program. Therefore, activating one thread only exposes all the parallelization-induced overhead without having any computational benefit at all. For this reason the one-thread runtimes for the independent and the augmented systems both result higher than the benchmark, sequential runtimes, which were compiled with optimization enabled but without active OpenMP\textsuperscript{\textregistered} flag.

Finally, the execution of the CUDA\textsuperscript{\textregistered} implementation\footnote{Only the results obtained for the compilation without the \texttt{--use-fast-math} options are shown. No significant runtime difference (less than 0.1 seconds was observed).} resulted the faster overall, taking $15.84$ seconds. The whole trajectory set was split into 10 stre-ams\footnote{By the result of the integer division of the 13509 states by 9, with the tenth stream containing a number of states equal to the remainder of that division.}. The stream definition guideline should in practice fit the application the propagator would run on, being the sole flexibility degree left by the implementation. However, from the GPU viewpoint, larger kernels always imply a better GPU exploitation, thus creating too many outer augmentation levels, i.e. activating a large number of independent CUDA\textsuperscript{\textregistered} streams, with too few trajectories each would result in a performance degradation, eventually obtaining what already observed with the independent integration cases for the single trajectories.

Table \ref{tab:finalcompare} summarizes the runtime results discussed in the previous lines for the different cases, for selected number of cores in the C implementation cases.
\begin{table}[h!]
	\centering 
	\def\svgwidth{\columnwidth}
	\caption{Runtimes for the independent runs and the augmented system executions. Average of 10 different runs each.}
	\begin{tabular}{c c c c }
		\textsc{\textbf{Case}} & \textsc{\textbf{Threads}} & \textsc{\textbf{GPU}} & \textsc{\textbf{Runtime [s]}} \\
		\hline
		C Independent & 1 & - & 245.02\\
		\hline
		C Augmented & 1 & - & 186.54\\
		\hline
		C Independent & 8 & - & 113.09\\
		\hline
		C Augmented & 8 & - & 40.60\\
		\hline
		C Independent & 40 & - & 113.54\\
		\hline
		C Augmented & 40 & - & 22.67\\
		\hline
		C Independent & 80 & - & 178.18\\
		\hline
		C Augmented & 80 & - & 18.70\\
		\hline
		CUDA\textsuperscript{\textregistered} Augmented & 4 & GTX 1050 & 15.84\\
		\hline
	\end{tabular}
	\label{tab:finalcompare}
\end{table}

\begin{figure}[h!]
	\tiny
	\centering 
	\def\svgwidth{\columnwidth}
	\includegraphics[width=\columnwidth]{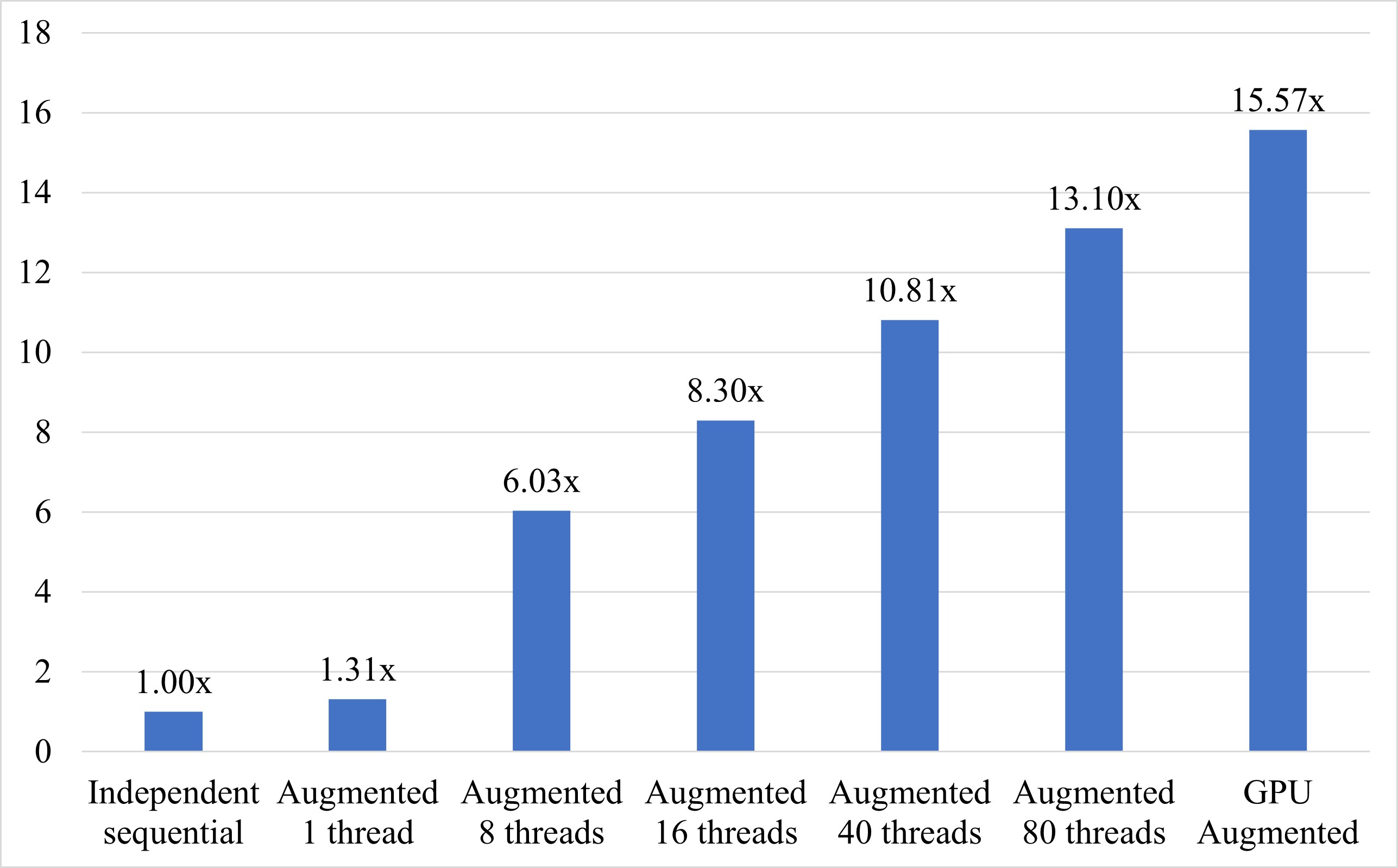}
	\caption{Speedup comparison among C and CUDA\textsuperscript{\textregistered} implementations.}
	\label{fig:histospeedup}
\end{figure}

Figure \ref{fig:histospeedup} shows the increasing speedup achieved by the augmented system when compared to the sequential and independent simulations of all the samples. The CUDA\textsuperscript{\textregistered} implementation runs more than 15 times faster compared to the baseline case, suggesting the suitability of the augmented PC algorithm to high performance and GPU computing facilities.

\subsection{CUDA\textsuperscript{\textregistered} Kernel profiling and optimization}
The final kernel implementations are the result of a long and detailed profiling and optimization process, performed with the Nsight\textsuperscript{\textregistered} Compute tool and a NVIDIA\textsuperscript{\textregistered} GeForce\textsuperscript{\textregistered} GTX 1050 graphics card. The choice of the 32 units block size of the dynamics kernel was driven mostly by the need of using the shared memory also for the temporary storage of the double precision state and the acceleration vector elements, just too large to be kept on the thread-private registers. The error computation kernels do not feature this bottleneck, thus the full amount of 1024 threads per block can be activated, minimizing the number of memory transactions and maximizing the reduction effects. The further four times of reduction while reading the error from ephemerides data (better detailed in Section \ref{subsec:error}) are the result of multiple trials: adding more reduction layers resulted in a kernel slowdown not compensated by the consequent speedup in the CPU function final call. Analogously, removing some of them resulted in a CPU function slowdown not compensated by the kernel speedup. The chosen number of while-reading reductions may however be affected by the problem size of the selected test case. Larger (or smaller) sets of initial conditions to be propagated may feature a different optimal implementation of the maximum error computation. On the contrary, 32 is already the minimum block size for the dynamics kernel and is not related to the problem size. Other non-relativistic dynamics function may drive the use of the shared memory in a different way, possibly allowing the use of larger block sizes.

The kernel performance is condensed into three numbers that result from the profiling process, i.e. the average multiprocessor occupancy for compute operations (\textsc{\textbf{Compute}} in Tables \ref{tab:profilefull} and \ref{tab:profiletenth}), the shared memory utilization (\textsc{\textbf{Shared Memory}} in Tables \ref{tab:profilefull} and \ref{tab:profiletenth}), and the memory throughput\footnote{Non-compute intensive but still highly parallelizable tasks may have the memory transfer as final bottleneck, this indicator measures how much of the CPU-GPU communication band width is used.} (\textsc{\textbf{Throughput}} in Tables \ref{tab:profilefull} and \ref{tab:profiletenth}). Despite profiling tools provide more detailed information, the presented indicators already allow the description of the kernel performances in sufficient detail. The profiling indicators of the cuBLAS\textsuperscript{\textregistered} kernels for matrix-matrix product (\texttt{dgemm}), matrix-vector product (\texttt{dgemv}), and element-wise summation (\texttt{daxpy})\footnote{For the sake of conciseness the kernels are identified with the names of the cuBLAS\textsuperscript{\textregistered} API functions, although optimized kernels are called for each specific GPU architecture and problem size.} used for the matrix multiplications embedded in the PC method are also shown, providing a performance comparison against well-known and heavily optimized library functions. In addition, also the kernel runtime as measured in the profiling activity is added to the comparison, to highlight the difference for the two tested augmented system sizes.

Table \ref{tab:profilefull} shows the kernel profiling results for an augmented system made of all the 13509 states to be propagated. The compute-bound kernels can be easily recognized as the ones exploiting the most the GPU's compute capability (Dynamics (\texttt{dynamics}), \texttt{dgemm}, error computation (\texttt{errCompute}), and maximum error reduction (\texttt{maxReduce})). The matrix-matrix multiplication is more memory bound than the dynamics function, because larger arrays must be loaded at the same time on the shared memory to perform the computation. On the contrary, the dynamics kernel features a more relevant computational bottleneck, because of the much more complex algorithm compared to the simple products and summations of the \texttt{dgemm} case. The remaining kernels, as well as the lower-intensity but still highly parallelizable error computation, all feature a high memory throughput. The GTX 1050 card used as a maximum memory band width of 112.1 Gb/s: the closer the throughput to that value the better the memory transactions are managed, essential feature for memory-bound problems. If some computations are added in the kernel, some throughput is inevitably lost, as latency sources are introduced between transaction to/from the shared/global memory.
\begin{table}[h!]
	\centering 
	\def\svgwidth{\columnwidth}
	\caption{13509-sized augmented state kernel profiling results.}
	\begin{tabular}{c c c c c}
		\textsc{\textbf{Kernel}} & \textsc{\textbf{Compute}} & \textsc{\textbf{Shared}} & \textsc{\textbf{Throughput}} & \textsc{\textbf{Runtime}} \\
				& [\%] & \textsc{\textbf{Memory}} [\%] & [Gb/s] & [$\mu$s] \\
		\hline
		\texttt{dynamics} & 97.11 & 24.96 & 1.91 & 137470 \\
		\hline
		\texttt{dgemm} & 94.15 & 26.67 & 4.06 & 164210 \\
		\hline
		\texttt{dgemv} & 47.14 & 84.54 & 93.24 & 1400 \\
		\hline
		\texttt{daxpy} & 18.18 & 74.78 & 60.26 & 32.48 \\
		\hline
		\texttt{errCompute} & 99.41 & 53.26 & 59.36 & 4740 \\
		\hline
		\texttt{maxReduce} & 36.80 & 71.36 & 79.25 & 279.36 \\
		\hline
	\end{tabular}
	\label{tab:profilefull}
\end{table}

Table \ref{tab:profiletenth} shows the kernel profiling results for an augmented system made of 1501 states to be propagated\footnote{This number results from splitting the full augmented system of 13509 states into 10 CUDA\textsuperscript{\textregistered} streams, each but the last with a number of states equal to the integer division between 13509 and 9. The tenth and last streams contains a number of states equal to the remainder of the previous integer division.}. The more compute-bound kernels do not show a significant loss of performance compared to the 13509-sized single augmented state case of Table \ref{tab:profilefull}, whereas the other kernels do. This highlights the suitability of GPU computing to extremely intense and parallelizable tasks, where the kernel call overhead is heavily compensated for by the massive task parallelization. Nevertheless, the capability of successfully processing also smaller-sized problems remains crucial for the program flexibility, particularly for what concerns the adoption of the proposed two-level augmentation scheme comprising smaller sub-systems significantly different from each other.
\begin{table}[h!]
	\centering 
	\def\svgwidth{\columnwidth}
	\caption{1501-sized augmented state kernel profiling results.}
	\begin{tabular}{c c c c c}
		\textsc{\textbf{Kernel}} & \textsc{\textbf{Compute}} & \textsc{\textbf{Shared}} & \textsc{\textbf{Throughput}} & \textsc{\textbf{Runtime}} \\
				& [\%] & \textsc{\textbf{Memory}} [\%] & [Gb/s] & [$\mu$s] \\
		\hline
		\texttt{dynamics} & 96.86 & 24.92 & 1.91 & 15300 \\
		\hline
		\texttt{dgemm} & 93.67 & 26.65 & 4.08 & 18390 \\
		\hline
		\texttt{dgemv} & 42.59 & 78.74 & 87.53 & 170.85 \\
		\hline
		\texttt{daxpy} & 9.19 & 33.95 & 36.31 & 6.62 \\
		\hline
		\texttt{errCompute} & 98.25 & 53.44 & 59.78 & 529.34 \\
		\hline
		\texttt{maxReduce} & 26.29 & 63.35 & 68.20 & 42.88 \\
		\hline
	\end{tabular}
	\label{tab:profiletenth}
\end{table}

Table \ref{tab:1vs10} compares instead the runtimes of the CUDA\textsuperscript{\textregistered} program, for the cases of 1 and 10 active streams. Theoretically, the former has the advantage of a better GPU resource exploitation, whereas the latter makes a more aggressive use of the CPU-GPU concurrency. Despite the lower GPU efficiency, the achieved runtimes are almost identical. Therefore, the two-level augmentation scheme can efficiently tackle the case of differently-sized lower-level augmented subsystems, showing the flexibility of the proposed computational scheme, despite the fine grain code optimization necessary for its successful implementation.
\begin{table}[h!]
	\centering 
	\def\svgwidth{\columnwidth}
	\caption{1 and 10 streams CUDA\textsuperscript{\textregistered} program executions. Average of 10 different runs each.}
	\begin{tabular}{c c}
		\textsc{\textbf{Case}} & \textsc{\textbf{Runtime}} \\
		& [s] \\
		\hline
		1 stream & 15.78 \\
		\hline
		10 streams & 15.84 \\
		\hline
	\end{tabular}
	\label{tab:1vs10}
\end{table}

\section{Conclusion}
This work explores the benefits that high-performance CPU clusters and GPU computing architectures bring to the short leg orbital propagation of large sets of initial conditions. The tested case studies the runs required to design a Solar Orbiter-like resonant phase with Venus, optimized to surf the relativistic N-body environment, proposing a two-level augmentation strategy implemented in the C and the CUDA\textsuperscript{\textregistered} programming languages.

Propagating the augmented system always outperforms propagating the trajectories independently, both in the sequential and the parallelized case. The augmentation benefits appear in two different aspects, the first being the reduced overhead compared to the repeated independent runs, the second represented by a finer grain parallelization also exploiting optimized linear algebra libraries.

The approach scalability allows its implementation on GPU architectures, with low end and old graphics card already capable of matching the performance of a common-size cluster node. Data center card models can make the algorithm run around 400-500 times faster, while the newest gaming GPUs should already allow a 60 times program acceleration. In addition, implementing the second order version of the modified PC method with error feedback should introduce a further two/three-fold acceleration.

Despite being shown on a test case derived by a trajectory optimization application, the proposed scheme represents a completely general orbital propagator. Any application requiring the propagation of large sets of initial conditions could benefit by the high computational efficiency provided by this algorithm, not necessarily requiring the use of supercomputing facilities in favour of much less expensive graphics cards.

Future works may keep developing the software tool, managing longer integration spans in sequence, and, as a general propagator, implementing flyby detection procedures. At the same time, the optimization scheme that led to the presented trajectory design application can be modified to accommodate massively parallel search approaches. On the implementation side, the benefits of using CUDA\textsuperscript{\textregistered} graphs instead of the stream-based management of the outer level augmented systems may also be explored.

\section*{Acknowledgments}
The research leading to these results has received funding from the European Research Council (ERC) under the European Union’s Horizon2020 research and innovation programme as part of project COMPASS (Grant agreement No 679086), \url{www.compass.polimi.it}, and the European Space Agency (ESA) through the Open Space Innovation Platform (OSIP) co-funded research project "Robust trajectory design accounting for generic evolving uncertainties", Contract No. 4000135476/21/NL/GLC/my.

\bibliography{Bibliography}

\end{document}